\documentclass[aps,prd,preprintnumbers,twocolumn,superscriptaddress,nofootinbib]{revtex4}
\usepackage{hyperref}
\hypersetup{colorlinks=true,linkcolor=purple,anchorcolor=blue,citecolor=blue, filecolor=blue,urlcolor=red,bookmarksnumbered=true,
	pdfview=FitB
}
\usepackage{color}
\usepackage{booktabs} 
\usepackage{threeparttable} 
\usepackage{xcolor}
\usepackage{ulem}
\usepackage{csquotes}
\colorlet{purple1}{blue!70!red}
\colorlet{darkred}{red!50!black}
\usepackage{slashed}
\usepackage{graphicx}
\usepackage[bf,SL,BF]{subfigure}
\usepackage{psfrag}
\usepackage{color}
\usepackage{amssymb}
\usepackage{amsmath}
\usepackage{epstopdf}
\usepackage{natbib}
\usepackage[mathlines]{lineno}

\def\ep{\epsilon}
\def\vep{\varepsilon}
\def\la{\langle}
\def\ra{\rangle}

\def\be{\begin{equation}}
\def\ee{\end{equation}}
\def\bea{\begin{eqnarray}}
\def\eea{\end{eqnarray}}

\begin{document}
	
	\title{Mixing effects on spectroscopy and partonic observables of mesons with logarithmic confining potential  in a light-front quark model}
 \author{Bhoomika Pandya}
 \email{bhumispandya@gmail.com} 
  \author{Bheemsehan Gurjar}
 \email{gbheem@iitk.ac.in} 
 \author{Dipankar Chakrabarti}
\email{dipankar@iitk.ac.in} 
\affiliation{Department of Physics, Indian Institute of Technology Kanpur, Kanpur 208016, India}
 \author{Ho-Meoyng Choi}
 \email{homyoung@knu.ac.kr}
 \affiliation{Department of Physics Education, Teachers College, Kyungpook National University, Daegu 41566, Korea}
 \author{Chueng-Ryong Ji}
 \email{crji@ncsu.edu} 
	\affiliation{Department of Physics, North Carolina state University, Raleigh, North Carolina 27695-8202, USA}
	
\begin{abstract}
Using the variational principle, we systematically investigate the mass spectra and wave functions of both $1S$ and $2S$ state heavy pseudoscalar $(P)$ 
and vector $(V)$ mesons within the light-front quark model. This approach incorporates a Coulomb plus logarithmic confinement potential to accurately describe the {constituent quark and antiquark} dynamics. Additionally, spin hyperfine interactions are introduced perturbatively to compute the masses of pseudoscalar and vector mesons.
The present analyses of the $1S$ and $2S$ states require the consideration of mixing between them to account for empirical constraints. 
These constraints include the mass gap $\Delta M_{P} > \Delta M_{V}$, where $\Delta M_{P(V)} = M^{2S}_{P(V)} - M^{1S}_{P(V)}$ and the hierarchy of the decay constants 
$f_{1S} > f_{2S}$. 
We find the optimal value of the mixing angle to be 
$\theta = 18^{\circ}$, significantly enhancing the consistency between our spectroscopic predictions and the experimental data compiled by the Particle Data Group (PDG).
Furthermore, based on
the predicted mass, the newly observed resonance $B_J(5840)$ could be {
assigned} as a $2^1S_0$ state in the $B$ meson family. 
The study also reports {
various} pertinent observables, {
including} twist-2 distribution amplitudes, electromagnetic form factors, 
charge radii, $\xi$ moments, and transition form factors which are found to be {
consistent with both} available lattice simulations and experimental data. 
{In addition, our predicted branching ratios for the channels of $B^+ \rightarrow \tau^+ \nu_{\tau}$ as well as rare decays of $B^0$ and $B_s^0$ appear in accordance with experimental data.}
\end{abstract}

\maketitle
\section{Introduction}
{Typical descriptions of various meson properties are encapsulated by either Euclidean or Minkowski approach. Calculations and simulations 
in {the Euclidean  approach typically involve Euclidean time for the analyses of two- and three-point correlation functions,} with source and sink operators specifying hadron states. The Euclidean
approach~\cite{Shifman19791,bali2018,asgarian2021,Dudek2006,padmanath2015,cooper2022} is employed not only in the Lattice Gauge Theories with
large time separation but also in the QCD Sum Rule (QCDSR)
calculations with operator product expansions. It provided {a} wealth of valuable insights into the behavior of quarks and gluons within hadrons. In the Minkowski approach, various {model-building} frameworks were developed to explore useful insights {into} the characteristics of hadrons. The static properties{, such as mass spectroscopy, were estimated by} traditional quark models employing specific Hamiltonians {that primarily incorporate} the Coulomb force at short {distances} and the confinement force at large {distances,} characterized by various potential terms. Such models have achieved remarkable success{,} notably in providing detailed descriptions of mesons~\cite{rosner2008,Ebert:2009ua,shah20141,Godfrey2016,pandya2021}. 
For the relativistic Hamiltonians, the light-front (LF) quantization has been utilized to incorporate factorization theorems of perturbative QCD alongside non-perturbative correlation functions such as parton distribution functions (PDFs) and distribution amplitudes (DAs) to examine hard inclusive and exclusive processes, respectively, in terms of the non-perturbative LF wave functions (LFWFs) of hadrons 
\cite{choi2009,ke2010,chang2018,hwang2012,Dhiman2019,adhikari2021,lan2019}.
The newer development with the holographic QCD describing hadrons as quantum fields propagating into bulk fields with extra dimensions has also been efficient in predicting mass
spectroscopy~\cite{karch2006,paula2009,de2009,branz2010,swarnkar2015,chang2017,Gurjar:2024wpq,Ahmady:2022dfv,Ahmady:2021yzh,Li:2022izo,Li:2021jqb,deTeramond:2021yyi}. The holographic platform implemented the confinement through the dilatation with the choice of soft vs. hard walls, incorporating quark-related fields via the infinite limit of $N_c$ and $N_f$.}

{In the present work, we utilize the light-front quark model (LFQM) based on the light-front dynamics (LFD) in the Minkowski approach to analyze both the meson spectroscopy and structures simultaneously. Within the LFQM formalism, the hadronic wave function comprises
equal LF time quantization naturally incorporating relativistic effects. Due to the rational energy-momentum dispersion relation in LFD, there is a sign correlation between the LF energy and LF longitudinal momentum. This feature provides the distinguished characteristic of the vacuum property in the LFD, sweeping the non-trivial vacuum condensation effects into the LF zero modes (LFZMs), i.e. the modes {with} zero LF longitudinal momentum, and leaving the rest of the vacuum clean by suppressing the quantum fluctuations of the non-LFZMs in the vacuum. Moreover, the longitudinal and transverse boost operators in LFD are kinematical and thus the LFWFs are completely boost invariant, i.e. independent of reference frames~\cite{brodsky1998,Choi1999}. 
Namely, the meson LFWFs obtained in the meson rest frame are identical to the corresponding LFWFs obtained in any other moving frames.    
These remarkable features make the LFQM a promising theory for the phenomenological study of hadron properties. Due to the simultaneous analyses of both spectroscopy and structures, 
one can analyze various physical observables and correlation functions including mass spectra, decay constants, form factors, DAs, PDFs, generalized parton distributions (GPDs), etc., all {within a single} designated LFQM.}

Early endeavors in the study of LFQM involved the successful utilization of the standard Cornell potential, where confinement {
exhibits} linear behaviour~\cite{Choi2015}. 
Subsequent efforts have investigated alternative approaches, including harmonic and exponential confinements~\cite{Dhiman2019}. Given {
that} previous studies on LFQM exclusively addressed ground states, the structure and features of radially excited states have not received much attention in these models. However, in a recent study ~\cite{Arifi2022}, two of us computed ground and radial excited states {
using a Coloumb} plus linear confinement potential in variational calculations, utilizing the two lowest-order harmonic oscillator basis functions. The study reports mass spectra for the 1S and 2S states both without and with consideration of mixing effects {
in} the wave function. 

The analysis of the first radial excited state {
is significant} in assessing two important empirical hierarchies observed from the experimental data. 
{
These} include the higher mass gap between pseudoscalar states compared to vector states, denoted as $\Delta M_{P} > \Delta M_{V}$, where $\Delta M_{P(V)} = M^{2S}_{P(V)} - M^{1S}_{P(V)}$. Additionally, it involves the larger decay constant of the ground state in comparison to the radial excited states, given as $f_{1S} > f_{2S}$. From this study, it is apparent that there is not much disparity between {
the masses} predicted by the pure and mixed treatment of wave functions~\cite{Arifi2022}. 
However, mixing effects are necessary for {
explaining} the aforementioned hierarchies. With the inclusion of mixing effects, the prediction made in Ref.~\cite{Arifi2022} has a $\chi^2$ analysis value of 0.009 with respect to the experimental data. \footnote{The $\chi^2$ values are computed using formula $\chi^2 = {\sum_i}[{(E_i-T_i)}^2/{T_i}^2$], where $E_i$ and $T_i$ represents data from PDG and theoretical predictions, respectively.} 
Also, the {
predictions from the exponential confinement  show a disparity compared to the experimental results, with a} $\chi^2$ analysis value of 0.012 for ground state heavy-light mesons. 

Except {
for} the exponential confinement, all prior attempts {
have used potentials that} belong to a special choice of the well-known general potential, $-Cr^{-\alpha}+Dr^{\beta}+V_0$ with $V_0 = a$, $\alpha = 1$ and $\beta = \nu$  ~\cite{Lucha1991,Patel2009}. However, the pure phenomenological potential, {
such as the} logarithmic potential $V(r)= c\ln(r/r_0)$ suggested by Quigg and Rosner \cite{Quigg1977,Rosner1979}, is believed to be successful in heavy-light and heavy-heavy meson spectroscopy. 
As far as the structural properties of mesons are concerned, the logarithmic potential indicates $n~| \psi(0)|^2$ is constant, where $n$ is the principle quantum number and $|\psi(0)|^2$ is the square of wave function at the origin. 
{This is in contrast to other potentials with power-law forms $V \sim r^\epsilon$ where $\epsilon > 0$, $|\psi(0)|^2 \sim n^{2(\epsilon-1)/(2+\epsilon)}$ \cite{Quigg1977,Rosner1979}. %
In this respect, the logarithmic potential exhibits a unique characteristic 
of producing level spacings independent of quark mass and flavour ~\cite{Quigg1977,Rosner1979}. This property contributes to the attractiveness of the logarithmic potential model in describing certain aspects of meson spectroscopy. Since the potential energy between the quark and antiquark varies logarithmically with their separation distance, the resulting energy spectrum tends to have {more uniform level spacings}, regardless of the quark masses or other characteristics.
This feature is particularly intriguing{, as the experimental data show that} the mass gap between the $1S$ and $2S$ states 
{is around 600 MeV, almost independent of the heavy flavors,} exhibiting some degree of universality. 
{This motivates us to comparatively analyze heavy flavor mesons using both} the logarithmic potential 
and the typical linear potential{,} including both $1S$ and $2S$ states.}

Hence, we aim to study the mass spectra and other related properties closely following the scheme developed in \cite{Choi:2007yu} but with the use of a logarithmic potential for mesons. 
Additionally, we seek to assess the impact of mixing effects on the observed experimental hierarchies. 
The $B$ meson spectrum has been enriched by the observation of novel states $B_J(5840)$ and $B_J(5960)$ by the LHC$_b$ Collaboration~\cite{LHCb2015}. 
Additionally, the new resonance labelled as $D_s(2590)$ {was} recorded by LHC$_b$ very recently~\cite{Lhcb2021ds}. Their experimental properties and mass range make them suitable candidate for the first radially excited state. However, their true identification is pending, 
and where to place them in the existing mass spectrum of $B$ and $D_s$ mesons 
remains an open problem. The present study of radially excited states could also 
aid in their identification. Along with the mass spectra, pseudoscalar decay constants cover a wide range of applicability including the di-leptonic decays of ground state pseudoscalar mesons in heavy-light sector. Estimation of these flavour changing neutral current transitions opens a window to possible existence of new physics.
The connection between di-leptonic decays and partonic observables in heavy-light mesons lies in their shared goal of probing the fundamental interactions and structure of hadrons as the  the momentum distribution amplitude for the constituent quark and antiquark inside the mesons are being measured immediately before their annihilation to lepton pairs. Thus, we also address 
well-known weak decays of heavy-light mesons as a complementary information from present study.

The remainder of the paper is structured as follows. In subsequent Sec. \ref{TF}, we outline the formalism which includes the inter quark-antiquark potential, effective Hamiltonian and trial wave functions employed in the present study. We also discuss the procedure of determining our model parameters following the variational principle within LFQM. In Sec. \ref{APP}, we summarize the explicit formulas to obtain the various physical quantities such as decay constants, DAs and electromagnetic form factors (EMFFs), transition form factors (TFFs) of quarkonia and di-leptonic decays of heavy-light mesons within LFQM. Our numerical results are presented and discussed in Sec. \ref{RD} followed by the summary in Sec. \ref{SUM}.
\section{Model Description}\label{TF}
{
The stationary meson system is characterized as an interacting bound-state consisting of valence quark and antiquark that are effectively dressed. 
This system conforms to the eigenvalue equation derived from the effective Hamiltonian motivated by QCD given by
\begin{equation}
H_{q\bar{q}}|\Psi_{q\bar{q}}\rangle = M_{q\bar{q}} |\Psi_{q\bar{q}}\rangle,
\end{equation}
where $M_{q\bar{q}}$ and $\Psi_{q\bar{q}}$ are the mass eigenvalue and eigenfunction of the $q{\bar q}$ meson bound state, respectively.
The effective Hamiltonian in the $q{\bar q}$ center of mass frame is represented as 
}
\begin{equation}\label{eq:Hqbarq}
H_{q\bar{q}} = H_0 + V_{q\bar{q}}{,}
\end{equation} 
where $H_{0}$ is the relativistic kinetic energy term given by 
\begin{equation}\label{eq:H0}
H_0 = {\sqrt{m_q^2 + {\mathbf{k}}^2}} + {\sqrt{{m_{\bar{q}}^2 + {\mathbf{k}}^2}}}
\end{equation} 
Here,  {$m_{q(\bar{q})}$ is the mass of the constituent quark (antiquark) with three momentum $\textbf{k}=({\bf k}_\perp, k_z)$.}
The usual quark-antiquark interaction potential $V_{q\bar{q}}$ is {typically} modeled as the sum of {a short-distance} Coulomb part $V_{\text{Coul}}$ 
and { a long-distance} confining part  $V_{\text{Conf}}$. 
{ Upon} inclusion of the spin-dependent hyperfine {interaction $V_{\text{Hyp}}$,  $V_{q\bar{q}}$ takes the following form}
\begin{equation}\label{eq:Vqbarq}
V_{q\bar{q}} = V_{\text{Coul}}\,+ V_{\text{Conf}}\,+V_{\text{Hyp}}{.}
\end{equation}
{The Coulomb potential is given by
\begin{align}\label{eq:coulpot}
V_{\text{Coul}}=-\frac{4 \alpha_s}{3 r},
\end{align}
where $\alpha_{s}$ {{is the strong coupling constant and considered in present scheme as a free parameter}, whereas 
$r$ is the distance between quark and antiquark. For the confining potential, we take
the logarithmic potential given by~\cite{Quigg1977,Rosner1979}}
\begin{align}\label{eq:confiningpot}
  V_{\text{Conf}}=  \,a + \, c\,\ln{(r/{r_0})},
\end{align} 
where $c$ and $r_0$ are the logarithmic confinement parameters adopted from~\cite{Quigg1977,Rosner1979}. 
Similarly, the hyperfine interaction potential{,} which is related to effective one-gluon exchanges and describes the spin-dependent part of the interaction between 
quarks within hadrons{,} is given by
\begin{align}
V_{\text{Hyp}}= {\frac{32 \pi \alpha_s}{9}} \frac{\left(\textbf{S}_{q}\cdot \textbf{S}_{\bar{q}} \right) {\delta^3\left(\textbf{r}\right)}} {m_q m_{\bar{q}}}{,}
\end{align}
{where $\langle\textbf{S}_{q}\cdot \textbf{S}_{\bar{q}} \rangle$ represents the spin-spin interaction contribution, having values of ${1/4}$  for spin triplet states 
and ${-3/4}$ for spin singlet states, respectively.
We  note that the two potential parameters $a$ and $\alpha_s$ are to be determined later.} \\
\begin{figure}[ht]
\centering
\includegraphics[width= 0.9\linewidth]{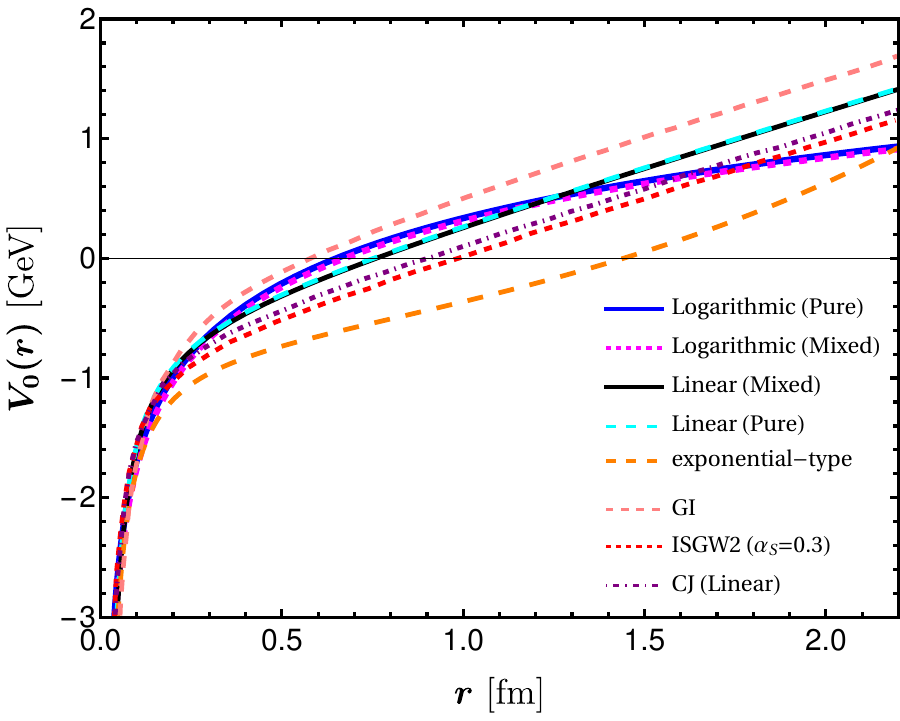}\hspace{0.2cm}
\caption{ Variation of central potential as a function of inter quark distance {$\mathbf{r}$ (in fm)}  for pure and mixed scenarios. The potential behaviour from other studies Linear~\cite{Arifi2022}, Exponential~\cite{Dhiman2019}, GI \cite{Godfrey1985},  ISGW2 \cite{Scora1995} and old CJ \cite{Choi20071} are also presented for comparison.}
\label{POT}
\end{figure}

{
For the four-momentum $P^{\mu}=\left(P^+,P^-,\textbf{P}_{\bot}\right)$ of a meson and the four-momentum $p^{\mu}_i$ of the $i$th constituent quark $(i=1,2)$,
the LFWF of the meson is represented by Lorentz invariant internal variables and helicity $\lambda_i$. The internal variables include
the longitudinal momentum fraction  $x_i = p_i^+/P^+$ and the relative transverse momentum $\textbf{k}_{{\bot}i} = \textbf{p}_{{\bot}i}-{x_i}\textbf{P}_{{\bot}}$ of the $i$th quark,
which satisfy $\sum_{i=1}^2\,x_i\,=\,1$ and $\sum_{i=1}^{2}\textbf{k}_{\perp i}\,=\,0$.
 We assign $i=1$  to the quark and $i=2$ to the antiquark, and define $x\equiv x_1$ and ${\bf k}_\perp\equiv {\bf k}_{\perp 1}$.
Then, the three-momentum ${\bf k}=(k_z, {\bf k}_\perp)$ can be written 
in terms of $(x,{\textbf{k}_\perp})$ as
$k_{\textit{z}} = \left(x-\frac{1}{2}\right)M_0 + \frac{{m_{\bar{q}}^2}-{m_{q}^2}}{2M_0}{,}$
where
\begin{equation}\label{eq:M0}
M_0^2 = \frac{\textbf{k}_{\bot}^2 + m_q^2}{x} + \frac{\textbf{k}_{\bot}^2 + m_{\bar{q}}^2}{1-x}
\end{equation}
is the boost-invariant meson mass squared. The Jacobian factor for the variable transformation $\{ k_z, {\bf k}_\perp\}\to \{x,{\textbf{k}_\perp}\}$ is given by
$\frac{\partial k_{\textit{z}}}{\partial x} = \frac{M_0}{4x(1-x)} \Big[1-\frac{(m_q^2-m_{\bar{q}}^2)^2}{M_0^4}\Big].$
}

The LFWF $\Psi_{q\bar{q}}=\Psi_{nS}^{J,J_z}$ for {the $n$S state meson} in momentum space is {then} given by~\cite{Choi:2007yu}
\begin{equation}\label{eq:8q}
{\Psi}_{nS}^{JJ_z}\left(x,\textbf{k}_{\bot},\lambda_i\right) = \Phi_{nS}\left(x,\textbf{k}_{\bot}\right) \mathcal{R}_{{\lambda_q}{\lambda_{\bar{q}}}}^{JJ_z}\left(x,\textbf{k}_{\bot}\right){,}
\end{equation}
{where} $\Phi_{nS}\left(x,\textbf{k}_{\bot}\right)$ and $\mathcal{R}_{{\lambda_q}{\lambda_{\bar{q}}}}^{JJ_z}\left(x,\textbf{k}_{\bot}\right)$ 
are the radial and spin-orbit wave functions{,} respectively. 
{The spin-orbit wave function} can be derived from the ordinary equal-time static one with assigned quantum number $J^{PC}$ through the interaction-independent Melosh
transformation. 
Essentially{,} it is more convenient to use the covariant form of $\mathcal{R}_{{\lambda_q}{\lambda_{\bar{q}}}}^{J{J_{z}}}$ for pseudoscalar and vector mesons as 
\cite{Jaus1990} 
\begin{equation}
\mathcal{R}_{{\lambda_q}{\lambda_{\bar{q}}}}^{00} =  \frac{-1}{\sqrt{2}{\widetilde{M}_0}} \ \bar{{u}}_{\lambda_q}(p_q) \  \gamma_5 \ {v}_{\lambda_{\bar{q}}} (p_{\bar{q}}){,}
\end{equation}
\begin{equation}
\mathcal{R}_{{\lambda_q}{\lambda_{\bar{q}}}}^{1J_z} = 
\frac{-1}{\sqrt{2}{\widetilde{M}_0}} \,{\bar{u}}_{\lambda_q}(p_q) 
\Big[\slashed{\epsilon}(J_z) - \frac{\epsilon \cdot (p_q-p_{\bar{q}})}{M_0+m_q+m_{\bar{q}}} \Big] \, v_{\lambda_{\bar{q}}}(p_{\bar{q}}),
\end{equation}
where $\widetilde{M}_0 \equiv \sqrt{M_0^2-(m_q-{m_{\bar{q}})^2}}$. Also, the transverse and longitudinal components of polarization vectors of vector mesons can be written as follows~\cite{Jaus1990}
\begin{align}\nonumber
\epsilon^{\mu}\left(\pm 1\right)  = &\left(0,\frac{2}{P^{+}} \epsilon_{\bot}(\pm){\cdot}\textbf{P}_{\bot}, \epsilon_{\bot}(\pm)\right){,} \\  
%
\epsilon^{\mu}\left(0\right) =& \frac{1}{M_0}\left(P^+,\frac{-M_0^2+{\textbf{P}_\bot^2}}{P^+},\textbf{P}_{\bot}\right){,}
\end{align}
with $\epsilon_{\bot}\left(\pm 1\right) = \mp \frac{1}{\sqrt{2}}\left(1,\pm i\right)$. 
{It is worth noting that the spin-orbit wave function satisfies
the unitary condition $\langle \mathcal{R}_{{\lambda_q}{\lambda_{\bar{q}}}}^{J J_z} |\mathcal{R}_{{\lambda_q}{\lambda_{\bar{q}}}}^{J J_z} \rangle = 1$ without any additional free parameters.

As in the case of Ref.~\cite{Arifi2022}, we shall analyze  the $1S$ and $2S$ states of pseudoscalar and vector mesons. 
We introduce mixing between the $1S$ and $2S$ radial wave functions, denoted as $\Phi_{nS}$ in Eq.~(\ref{eq:8q}), 
by allowing a combination of the two lowest-order harmonic oscillator (HO) wave functions, $\phi_{1S}$ and $\phi_{2S}$, through coefficients $c_{nm}$ as follows
\begin{equation}\label{PhinS}
\Phi_{nS}= \sum^{2}_{m=1} c_{nm}\phi_{mS},
\end{equation}
where 
\begin{eqnarray}\label{eq:phi1S}
	\phi_{1S} (x, \mathbf{k}_\bot) &=& \frac{4\pi^{3/4}}{ \beta^{3/2}} 
	\sqrt{\frac{\partial k_z}{\partial x}} e^{-{\bf k}^2/ 2\beta^2},
	\nonumber\\
	\phi_{2S} (x, \mathbf{k}_\bot) &=& \frac{4\pi^{3/4}}{ \sqrt{6}\beta^{7/2}} 
	\left( 2 {\bf k}^2 - 3\beta^2 \right) \sqrt{\frac{\partial k_z}{\partial x}} e^{-{\bf k}^2/ 2\beta^2},
\nonumber\\
\end{eqnarray}
and $\beta$ is the variational parameter to be determined from our mass spectroscopic analysis. 
The HO basis wave functions $\phi_{nS}$ defined in Eq.~\eqref{eq:phi1S} satisfy the following normalization
\begin{equation}
\langle \phi_{nS}|\phi_{nS} \rangle=\int_{0}^{1} \,dx \int \frac{d^2\textbf{k}_\bot}{16\pi^3} |\phi_{nS}\left(x,\textbf{k}_{\bot}\right)|^{2} = 1.
\end{equation} 
The mixed states $\Phi_{nS}$ also adhere to the orthonormal condition, $\langle \Phi_{nS}|\Phi_{mS}\rangle=\delta_{nm}$.
Consequently, we express the expansion coefficients $c_{nm}$ in terms of the mixing angle $\theta$ between the pure states $\phi_{1S}$ and $\phi_{2S}$ as  
$c_{11}=c_{22}=\cos\theta$ and $c_{12}=-c_{21}=\sin\theta$.

We now compute the mass eigenvalue of the meson as $M^{nS}_{q{\bar q}}=\langle\Psi_{q{\bar q}}|H_{q{\bar q}}|\Psi_{q{\bar q}}\rangle=\langle\Phi_{nS}|H_{q{\bar q}}|\Phi_{nS}\rangle$.
The analytical expressions of $M^{nS}_{q{\bar q}}(n=1,2)$ for the mixed $(1S, 2S)$ state mesons are then given as 
\begin{widetext}
\begin{eqnarray}\label{eq:mass_eigen_value}
	M_{q\bar{q}}^{1S} &=& \frac{\beta}{ \sqrt{\pi}} \sum_{i=q,\bar{q}} \biggl\{ z_i e^{z_i/2} \biggl[  \frac{1}{3} c_{12}^2 (3-z_i) z_i~K_2\left(\frac{z_i}{2}\right)   
	+\frac{1}{6}  \left(9 - 3c_{11}^2 + 2c_{12}^2z_i^2 -6\sqrt{6}c_{11} c_{12}\right)K_1\left(\frac{z_i}{2}\right)  \biggr] 
	\nonumber\\ & & \mbox{} \qquad \qquad
	+ \sqrt{\pi} \left(\sqrt{6}c_{11} c_{12} - 3c_{12}^2\right) U\left(-1/2,-2,z_i\right)\biggr\}	 
	\nonumber \\ && \mbox{} 
	+ a+c\Bigg(1-\frac{\gamma_{E} }{2}-\log (2\beta r_{0})-\frac{2  c_{11}c_{12} }{\sqrt{6}}+\frac{ c_{12}^2}{3}\Bigg)
	- \frac{4\alpha_s \beta}{9\sqrt{\pi}} \left( 5+c_{11}^2 + 6\sqrt{\frac{2}{3}}c_{11} c_{12} \right)  
	\nonumber\\ && \mbox{} 
	+  \frac{16\alpha_s\beta^3 \langle\mathbf{S}_q\cdot \mathbf{S}_{\bar{q}}\rangle}{9m_q m_{\bar{q}} \sqrt{\pi}} 
	(3-c_{11}^2 + 2\sqrt{6}c_{11} c_{12}),
	\nonumber\\
	M_{q\bar{q}}^{2S} &=& M_{q\bar{q}}^{1S} (c_{11}\to c_{21}, c_{12}\to c_{22})
\end{eqnarray}
\end{widetext}
where $z_i=m_i^2/\beta^2$, $K_n$ is the modified Bessel function of the second kind of order $n$, $U(a,b,z)$ is confluent hypergeometric function and $\gamma_E$ is the Euler gamma function.
{We should note that the mass eigenvalues for the pure ($1S$, $2S$) states can be obtained by setting $\theta=0$, i.e.,
$c_{11}=c_{22}=1$ and $c_{12}=c_{21}=0$ in Eq.~\eqref{eq:mass_eigen_value}.

As discussed in Ref.~\cite{Arifi2022}, the optimal value of the mixing angle $\theta$  can be constrained by the experimentally observed mass gap relation 
between the $1S$ and $2S$ state heavy pseudoscalar and vector mesons, 
i.e. $\Delta M_{P} > \Delta M_{V}$, where $\Delta M_{P(V)} = M^{2S}_{P(V)} - M^{1S}_{P(V)}$. 
Moreover, because both pseudoscalar and vector mesons with the same $q{\bar q}$ contents share common $\beta$ parameters, 
as illustrated in Eq.~\eqref{eq:mass_eigen_value}, the mass gap is exclusively determined by the hyperfine interaction $V_{\rm Hyp}$.  Its explicit form
is given by~\cite{Arifi2022}
\begin{eqnarray} \label{eq:mass_gap}
\Delta M_P - \Delta M_V &=& C \left(2 \sqrt{6}\sin 2\theta - \cos 2\theta \right),
\end{eqnarray}
where $C=16  \alpha_s \beta^3 / (9m_q m_{\bar{q}} \sqrt{\pi})$. 
It is apparent that the pure $(1S, 2S)$ states without mixing (i.e., $\theta=0^\circ$)  always leads to  $\Delta M_{P} < \Delta M_{V}$,
necessitating the introduction of the mixing scheme in the present study. The condition $\Delta M_{P} >\Delta M_{V}$ leads to the following constraint on the mixing 
angle,  $\frac{1}{2}\, \cot^{-1} (2\sqrt{6})\simeq 6^\circ < \theta < 45^\circ$~\cite{Arifi2022}.   

To extract the model parameters in our study, 
we closely adhere to the methodology outlined in~\cite{Arifi2022,Choi2015}. 
Specifically, we utilize the radial wave functions $\Phi_{nS}(x, {\bf k}_\perp)$ as trial functions for the variational principle applied to the QCD-motivated
Hamiltonian $H_{q{\bar q}}$. The minimization of the mass eigenvalues is achieved by setting the derivative of the expectation value of the central Hamiltonian 
with respect to $\beta$ to zero: $\frac{\partial \langle\Phi| H_0 + V_{\rm Coul} + V_{\rm Conf}|\Phi\rangle}{\partial \beta} = 0$. 
We treat $\langle\Phi| V_{\rm Hyp}|\Phi\rangle$ as perturbation, ensuring common $\beta$ values for both pseudoscalar and vector mesons of the same $q{\bar q}$ content.
Subsequently, we determine the optimal values of  the mixing angle and the model parameters, including
the quark masses $(m_q,m_s,m_c,m_b)$, the potential parameters $(\alpha_s,a)$ and the variational parameters $(\beta_{q\bar{q}})$ for each meson. 

The variational principle permits writing $\alpha_s$ in terms of the other parameters, thereby reducing the degree of freedom in parameter space~\cite{Choi2015}. 
Consequently, the condition $\alpha_s = \alpha_s(a, m_q, m_{\bar{q}},\theta, \beta_{q\bar{q}})$ holds, allowing one to express $\alpha_s$ in terms of other parameters~\cite{Choi2015}.}
For the present study, we adopt the values of $c$ = 0.733 GeV and $r_0$ = 0.89 GeV$^{-1}$ suggested by Quigg and Rosner~\cite{Rosner1979}. 
Through {
an iterative analysis, we consider  quark masses of} $m_q$ = 0.22 GeV, $m_s$ = 0.45 GeV, $m_c$ = 1.68 GeV and $m_b$ = 5.10 GeV. 
{
These values align with those presented in~\cite{Arifi2022}.}
The remaining parameters, $a$ and $\alpha_s$, can be {
determined using} the two experimental masses of the $1S$ state as input. 
After {
exploring various} combinations of $(1^3S_1, 1^1S_0)$, 
we found that {
using} experimentally measured values of $B_s(1^3S_1)$ and $B_s(1^1S_0)$ as inputs 
can {
yield other meson} ground state masses close to the PDG values. Once the constituent quark masses, $a$ and $\alpha_s$ are determined, variational parameters  for each system will automatically {
be} deduced. 
{
It is important to note that, in the present scheme, the confining and short-range Coulomb parts are considered to be scale and flavor independent, as suggested in Ref.~\cite{Arifi2022, Choi2015}. Consequently, the numerical values of $a$ and $\alpha_s$ are the same for all the mesons under consideration. 
In this context, the quantity $\alpha_s$ ceases to be a strong coupling constant and can be regarded as a free parameter to be optimized, as discussed earlier.}
\begin{table*}
\begin{center}
\tabcolsep 5pt
\small
\caption{The constituent quark masses, potential parameters $a$ and $\alpha_s$ along with variational parameters $\beta_{q\bar{q}}$ for pure and mixed case. All are in units of GeV except the $\alpha_s$ which is dimensionless. Here we consider $q$ = $u,d$.}
\label{input1}
\begin{tabular}{lccccccccccccc}
\hline \hline
 &  $m_q$ & $m_s$ & $m_c$ & $m_b$ & a & $\alpha_s$ & $\beta_{bb}$ & $\beta_{bc}$ & $\beta_{bs}$ & $\beta_{bq}$ & $\beta_{cc}$ & $\beta_{cs}$ & $\beta_{cq}$\\
\hline
Pure & 0.22 & 0.45 & 1.68 & 5.10 & -0.9017 & 0.1422 &  1.3730 & 0.9952 & 0.7384 & 0.6863 & 0.7839 & 0.6056 & 0.5639\\
Mixed $\left( \theta= 18^\circ\right)$& 0.22 & 0.45 & 1.68 & 5.10 & -0.9211 & 0.1591 &  1.0939 & 0.7882 & 0.5820 & 0.5407 & 0.6193 & 0.4763 & 0.4435\\
\hline \hline
\label{Tab:variational_parameters}
\end{tabular}
\end{center}
\end{table*}

{
We obtain the optimal model parameters for both the pure state and the mixed state cases, respectively. 
In the mixed state case, we find the optimal mixing angle to be $\theta=18^\circ$, which is larger than the predicted value of $\theta=12^\circ$ in
the case of the linear confining potential~\cite{Arifi2022}.
Our results for the optimal model parameters for both pure and mixed states are summarized in Table~\ref{Tab:variational_parameters}.
}

{{Figure~\ref{POT} shows the central potential $V_0 = V_{\text{Coul}} + V_{\text{Conf}}$ as a function of radial distance between quark and anti-quark for pure $\left( \theta= 0^\circ\right)$ and mixed configurations  $\left( \theta= 18^\circ\right)$. Since the numerical values of two potential parameters $a$ and $\alpha_s$ exhibits close proximity in both cases (see Table~\ref{input1}), the behaviour of the potential from the pure and mixed scenarios tends to be similar. The quantitative trend of linear raising is also analogous to linear confinement potential with pure $\left( \theta= 0^\circ\right)$ and mixed configurations $\left( \theta= 12^\circ\right)$~\cite{Arifi2022}, old CJ \cite{Choi20071} as well exponential confinement \cite{Dhiman2019} within the LFQM up to the region over which the potential can be considered to be tested. However, the slight difference of the central potential may leads to large impact on the predictions of mass and other physical quantities. The central potential behavior from well-known models such as the Godfrey and Isgur (GI) model~\cite{Godfrey1985}, as well as the simple quark model known as the Isgur-Scora-Grinstein-Wise (ISGW2) model ~\cite{Scora1995} by Scora and Isgur is also illustrated for comparison.}}

\begin{figure*}
\centering
\includegraphics[width=\linewidth]{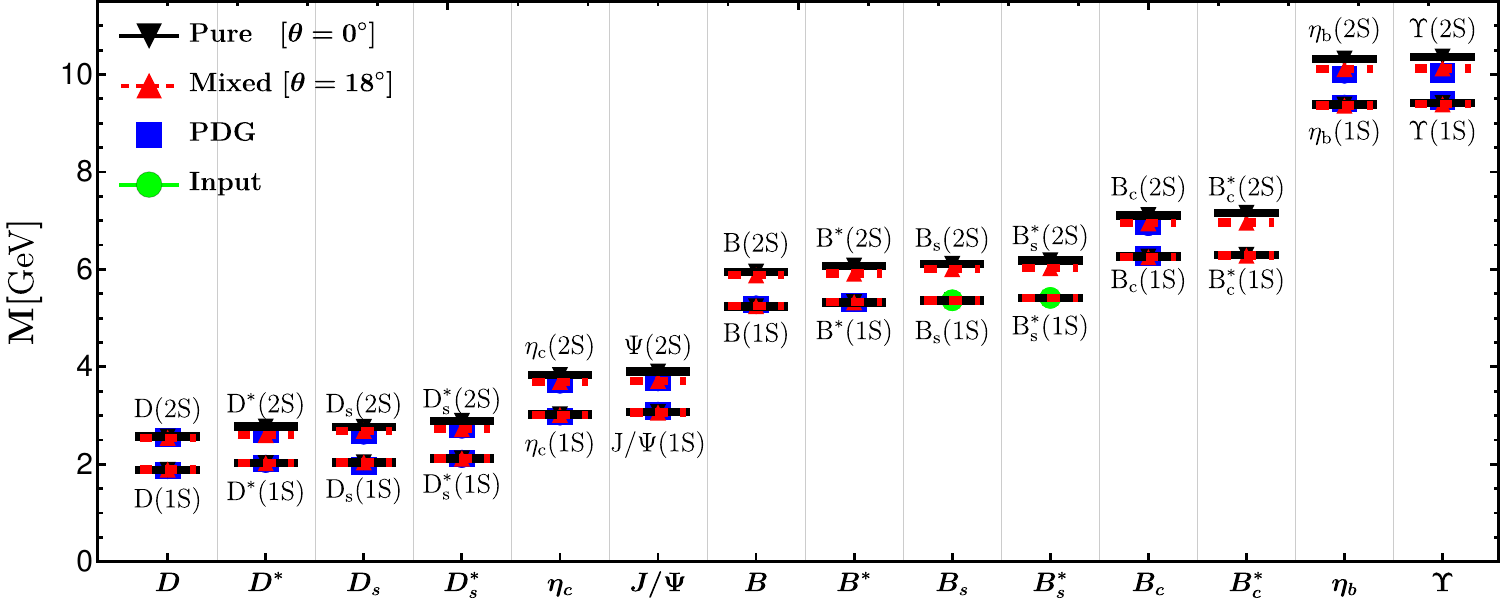}
\caption{The mass spectra of the ground $(1S)$ and first radial excited $(2S)$ states of the mesons in pure ($\theta=0^{\circ}$) and mixed ($\theta=18^{\circ}$) configurations shown as solid-black and red-dashed lines, respectively. The comparison is made with experimental data from PDG represented by blue squares~\cite{PDG2022}.}
\label{Fig:mass spectroscopy1}
\end{figure*}

\section{Applications}\label{APP}
In this section, we provide a concise overview of the calculations involved in determining additional observables to assess the predictivity of the present study. These observables include pseudoscalar and vector decay constants and DAs, electromagnetic form factors, charge radii, 
transition form factors of quarkonia, as well as di-leptonic decays of heavy-light mesons. 

\subsection{{Decay Constants and Distribution Amplitudes}}
The precise determination of the pseudoscalar and vector decay {constants} is crucial as {they} encapsulate the internal structure of {hadrons}. 
The pseudoscalar {meson} decay constant $(f_P)$ and vector {meson} decay constant $(f_V)$ can be expressed in terms of the parametrization 
of the weak current matrix elements of {the meson and the} vacuum. 
{Accordingly, the decay constants for pseudoscalar and vector mesons are defined as follows~\cite{Ball:1998sk,Ahmady:2019hag,Gurjar:2024wpq}
\begin{equation}
\left<0|\bar{q}(0)\gamma^{\mu} \gamma_{5} {q}(0)|P \right> = i f_P \ P^\mu,
\end{equation}
for pseudoscalar meson and 
\begin{equation}
\left<0|\bar{q}(0)\gamma^{\mu} {q}(0)| V(P,\lambda) \right> =  f_V \ M_V~\epsilon_{\lambda}^{\mu},
\end{equation}
for a vector meson with longitudinal $(\lambda = 0)$ polarization.
Here, $q(\bar{q})$ represent the field operators for quark (anti-quark) located at {the} same space-time points and 
$M_{V}$ denotes the mass of the vector meson.

The calculation of decay constants can be facilitated by employing the plus component $(\mu=+)$ of the current. The explicit expressions for the decay constants 
of the $nS$ state pseudoscalar and vector mesons are given by~\cite{Choi2021,Choi2014,Choi:2007yu}.
\begin{equation}\label{DCs}
f^{nS}_{{P}{(V)}} =2 \sqrt{6}\int_{0}^{1} dx\int \frac{d^2\textbf{k}_\bot}{16\pi^{3}} \frac{\Phi_{nS}\left(x,\textbf{k}_{\bot}\right)}{\sqrt{\mathcal{A}^2 + {\textbf{k}^2_\bot}}} \mathcal{O}_{P(V)}, \,
\end{equation}
where $\mathcal{O}_P = \mathcal{A}$ and $\mathcal{O}_V$ = 
$\mathcal{A}$ + $\frac{2{\textbf{k}^2_\bot}}{\textbf{D}_{LF}}$ with definitions $\mathcal{A} = (1-x) m_q + x \, m_{\bar{q}}$ and $\textbf{D}_{LF}$ = $M_0 + m_q + m_{\bar{q}}$. 
Indeed, the decay constants have been demonstrated to be independent of the current component and polarization; further detailed analysis on this topic can be found in~\cite{Choi2021,Choi2014}.

In this work, although the distribution amplitudes derived from various components of the currents and polarization vectors result in distinct classifications by twists, 
our analysis will specifically concentrate on the twist-2 DAs.
The twist-2 quark DAs can be obtained from Eq.~\eqref{DCs} as \cite{Choi:2007yu}
\begin{equation}
\phi^{nS}_{P(V)}(x)=\frac{2\sqrt{6}}{f^{nS}_{P(V)}}\int \frac{d^2\textbf{k}_\bot}{(2\pi)^{3}} \frac{\Phi_{nS}\left(x,\textbf{k}_{\bot}\right)}{\sqrt{\mathcal{A}^2 + {\textbf{k}^2_\bot}}} \mathcal{O}_{P(V)},\,
\end{equation} 
so that it can be normalized as 
\begin{eqnarray}\int_{0}^{1} {{\phi}}^{nS}_{P(V)}\, (x)\, dx = 1.
\end{eqnarray}
The expectation value of the
longitudinal momentum, given by $\xi\,=\,x-\,(1-x)\,=\,2x-1$ and known as $\xi$-moments is expressed as follows~\cite{Choi:2007yu}
\begin{eqnarray}
\langle \xi^m \rangle_{nS}  = \int_{0}^{1} \mathrm{d}x\xi^m {\phi}^{nS}_{P(V)}\, (x).\,
\end{eqnarray}

\subsection{Electromagnetic and Transition Form Factors}
The electromagnetic form factor of a pseudoscalar meson is calculated in the Drell-Yan-West frame where
$q^+\,= q^0 + q^3=0$ with $\textbf{q}^{2}_\bot \, = Q^2\, = -q^2$. 
In this frame, using the $+$ component of the current, 
the form factor can be derived as $F_{\rm em}(Q^2)=\frac{\langle P'|J^+|P \rangle}{2P^+}$ with the explicit form provided by \cite{Choi1999}
\begin{eqnarray}\nonumber
F^{nS}_{\rm em}(Q^2) &=& e_q \int_{0}^{1} dx\int \frac{d^2\textbf{k}_\bot}{16\pi^3} \Phi_{nS}\left(x,\textbf{k}_{\bot}\right) \Phi^{*}_{nS}\left(x,\textbf{k}'_{\bot}\right) 
\\ &&  \times \frac{\mathcal{A}^2 + \,{\textbf{k}_\bot}\cdot{\textbf{k}'_\bot}}{\sqrt{\mathcal{A}^2 + {\textbf{k}^2_\bot}}\sqrt{\mathcal{A}^2 + {\textbf{k}^{'2}_\bot}}}
+ e_{\bar q}(m_q \leftrightarrow m_{\bar{q}}),
\nonumber\\
\end{eqnarray} 
where $e_q (e_{\bar{q}})$ is the electric charge of quark (anti-quark) and
$\textbf{k}'_\bot = {\textbf{k}_\bot} + \left(1-x\right) {\textbf{q}_\bot}$. 
The EMFF is normalized such that $F_{\rm em}\left(0\right)\,=\,e_q\,+\,e_{\bar{q}}$ and the charge radius is computed as
\begin{equation}
\langle r^2 \rangle_{nS}  = -6 \,\frac{dF^{nS}_{\rm em}(Q^2)}{dQ^2}\bigg\rvert_{Q^{2}\,=\,0}.
\end{equation}
The transition form factors $F_{P\gamma}$ for the $P\to\gamma^*\gamma$ transition can be derived from
the matrix element of electromagnetic current as 
\be\label{Eq1}
\la\gamma(P-q)|J^\mu|P(P)\ra = i e^2 F_{P\gamma}(Q^2)\ep^{\mu\nu\rho\sigma}P_\nu\vep_\rho q_\sigma,
\ee
where $P^\mu$ and $q^\mu$ are the four momenta of the incident pseudoscalar meson and virtual photon,
respectively and $\vep$ is the transverse polarization vector of the final (on-shell)
photon. 

In Refs.~\cite{Ryu:2018egt,Choi:2017pi}, the TFFs were obtained from both $q^+=0$ frame and $q^+=P^+$ frame and were demonstrated to be
independent of the choice of reference frame. Moreover, the authors showed that the result obtained from the $q^+=P^+$ frame exhibit a salient feature 
that distinguishes it from those derived from the $q^+=0$ frame. This notable advantage of the $q^+=P^+$ frame not only 
confirms the boost invariance of the results but also facilitates a more effective computation of the timelike form factor compared to the commonly used $q^+=0$ frame.

The explicit form of the TFFs for the $nS$ state heavy quarkonina $\eta_{c(b)}\to\gamma^*\gamma$ transition in the $q^+=P^+$ frame is provided by~\cite{Ryu:2018egt,Choi:2017pi}
\begin{align}\label{eq:TFFsLFQM}\nonumber
   F^{nS}_{\eta_{c(b)}\gamma}(q^{2})=&e_{c(b)}^{2}\frac{\sqrt{2N_{c}}}{4\pi^{3}} \int_{0}^{1}\frac{{\rm d}x}{(1-x)}\int{\rm d}^{2}{\bf{k}_{\perp}}\frac{1}{M_{0}^{2} - q^{2}}\\
   &\times \Psi^{nS}_{\frac{\uparrow\downarrow-\downarrow\uparrow}{\sqrt{2}}}(x,\bf{k}_{\perp}),
\end{align}
with $N_c=3$ is the number of colors.  The LFWF  $\Psi_{\frac{\uparrow\downarrow-\downarrow\uparrow}{\sqrt{2}}}(x,\bf{k}_{\perp})$ of a heavy quarkonia where quark and antiquark masses are equal $m_{Q}=m_{\bar{Q}}$ is defined as~\cite{Ryu:2018egt,Choi:2017pi}
\begin{align}
   \Psi^{nS}_{\frac{\uparrow\downarrow-\downarrow\uparrow}{\sqrt{2}}}(x,\mathbf{k}_{\perp})=\frac{m_{Q}}{\sqrt{\mathbf{k}_{\perp}^{2}+m_{Q}^{2}}}  \Phi_{nS}(x,\mathbf{k}_{\perp}).
\end{align}
We should note that the direct calculation of the timelike TFF  is particularly effective due to the the simple pole structure $(M^2_0 - q^2)^{-1}$ in the timelike region ($q^2 >0$). 
By performing analytic continuation from timelike  region ($q^2>0$) to spacelike region ($q^2\to -q^2$),
the TFF can be determined in the spacelike region ($Q^2=-q^2$) without singularities.
}

\subsection{Di-leptonic Decays}
The charged pseudoscalar mesons $B^{+}$, $D^{+}$ and $D_s^{+}$ can undergo annihilation through a $W^{\pm}$ boson into a lepton-neutrino pair $(l^{+}\nu_l)$. 
The presence of highly energetic leptons in the final states makes these decays experimentally prominent, while the absence of hadrons in the final 
state renders these decays theoretically clean to predict, depending only on a single hadronic parameter: the pseudoscalar decay constant~\cite{villa2007}. 
Assuming that the main contribution to these weak leptonic decay transitions comes from the virtual boson-mediated annhilation of the bound quark-antiquark pair 
$(q_1 \bar{q_2})$ inside the pseudoscalar meson $\mathcal{P}$, the partial decay width is given by ~\cite{villa2007,PDG2022,rosner2008}
\begin{align}
\Gamma(\mathcal{P}^+ \rightarrow l^+\nu_l)=\frac{G_F^2}{8\pi} f^2_{\mathcal{P}} |V_{{q_1}{q_2}}|^2 m_l^2 \left(1-\frac{m_l^2}{M^2_{\mathcal{P}}}\right)^2 M_{\mathcal{P}}\,,
\end{align}
where $M_\mathcal{P}$ and $f_{\mathcal{P}}$ denote the mass and the decay constant of a ground state pseudoscalar meson, respectively. Our predicted values of $M_\mathcal{P}$ and $f_{\mathcal{P}}$ are being used in the computation of these transitions.
The values of other parameters such as Fermi's coupling constant $G_F = 1.16 \times 10^{-5}$ GeV$^{-2}$, mass of leptons as $m_e = 0.0005$ GeV, $m_{\mu} = 0.105$ GeV, $m_{\tau} = 1.776$ GeV,
CKM matrix elements 
$V_{ub} = (3.8 \pm 0.20) \times 10^{-3}$, $V_{cd} = 0.221 \pm 0.004$ and $V_{cs} = 0.975 \pm 0.006$ are taken from PDG~\cite{PDG2022}. The branching ratio for these transitions is then obtained as    
\begin{equation}
\mathcal{B} = \Gamma_{(\mathcal{P^+} \rightarrow l^+ \nu_l)} \times \tau_{\mathcal{P}}
\end{equation}
where $\tau_{\mathcal{P}}$ is lifetime of the respective mesons. We take $\tau_{D^+}$ = $1.033 \times 10^{-12}$ s, $\tau_{D_s^+}$ = $0.501 \times 10^{-12}$ s and $\tau_{B^+}$ = $1.638 \times 10^{-12}$ s as listed in PDG \cite{PDG2022}. 

The di-leptonic decay of neutral $B_q^0$ mesons into two charged lepton pairs $(l^+l^-)$ is suppressed by Glashow-Iliopoulous-Maiani (GIM) mechanisms and helicity constraints. 
However, such transitions occur through higher-order diagrams involving flavour-changing neutral currents. The presence of higher-order interaction vertices significantly reduces its decay probability. Thus, this class of di-leptonic decays are known as rare decays. The decay width for these transitions in neutral charge mesons is expressed as
\cite{bobeth2014,bobeth20142,buchalla1993}
\begin{eqnarray}\label{raredecay}
\Gamma_{(B^0_q \rightarrow \ell^+ \ell^-)} &=& \frac{G_F^2}{\pi} \frac{\alpha^2 f^2_{B_q} m_{\ell}^2}{(4 \pi \sin^2\Theta_W)^2 } \ m_{B_q} \\ && \nonumber  \sqrt{1 - 4 \frac{m_{\ell}^2}{m_{B_q}^2}} |V_{tb}^* V_{tq}|^2 |C_{10}|^2 
\end{eqnarray}
where the Weinberg angle is approximated as $\Theta_W(\approx 28^\circ)$ \cite{lee2015}.  
The PDG values for CKM matrix elements  $V_{tb}$, $V_{td}$ and $V_{ts}$ as $1.014 \pm 0.029 $, $ (8.6 \pm 0.2)\times 10^{-3}$ and $(41.5 \pm 0.9)\times 10^{-3}$
respectively are considered~\cite{PDG2022}. Within the Standard Model, rare decays are predominantly governed by the operator $\mathcal{O}_{10}$ with the corresponding Wilson coefficient $C_{10}$ given as \cite{buchalla1993,buras1998}
\begin{equation}
C_{10} = \eta_Y \frac{x_t}{8}\left[ \frac{x_t-4}{x_t-1}+\frac{3x_t}{(x_t-1)^2} \ln x_t \right]
\end{equation}
with  $x_t = (m_t / m_W)^2$; $m_t$ = 172.57 GeV and $m_W$ = 80.36 GeV~\cite{PDG2022}. The $\eta_Y = 1.026$ is the next leading order corrections \cite{buras1998}. 
The branching ratio for rare transitions is given by
\begin{equation}
\mathcal{B} = \Gamma_{(B^0_q \rightarrow \ell^+ \ell^-)} \times \tau_{B_q^0}
\end{equation}
We take experimental values $\tau_{B^0}$ = $ (1.517 \pm 4) \times 10^{-12}$ s and $\tau_{B_s^0}$ = $(1.527 \pm 0.0011) \times 10^{-12}$ s~\cite{PDG2022}.

\begin{figure*}
\includegraphics[width=\textwidth]{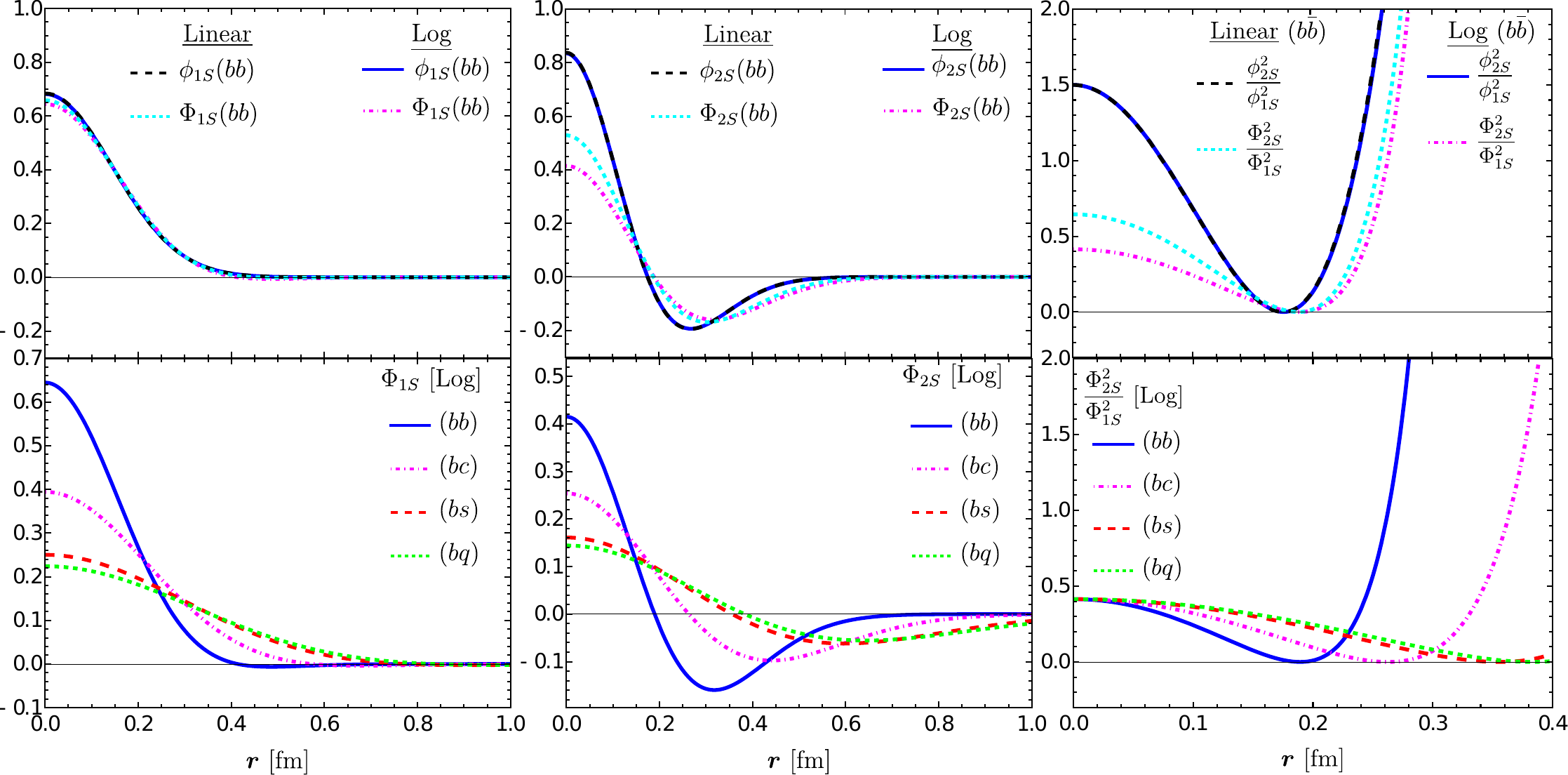}
\caption{Upper Panel: Comparison of the pure ($\phi$) and mixed ($\Phi$) radial wave functions for $1S$ and $2S$ states of bottomonium  ($b\bar{b}$) from present study and those obtained by linear confining potential~\cite{Arifi2022}.
It is noticed that $2S$ wave function has sizable mixing effect whereas $1S$ wave function has marginal modification. Lower Panel: The radial wave functions $\Phi_{1S}$ and $\Phi_{2S}$ for the various bottom flavoured mesons for mixed case.}
\label{Fig:LFWFs}
\end{figure*}
\begin{figure*}\label{contri}
\centering
\includegraphics[width=\linewidth]{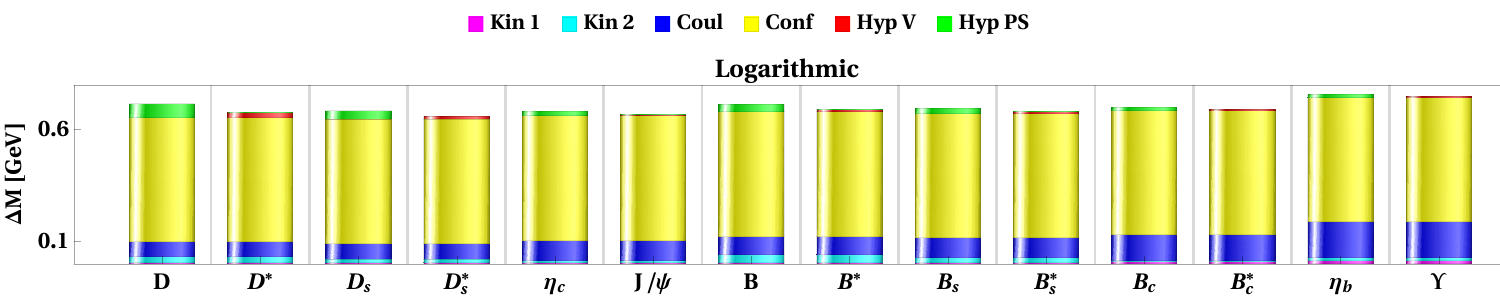}
\includegraphics[width=\linewidth]{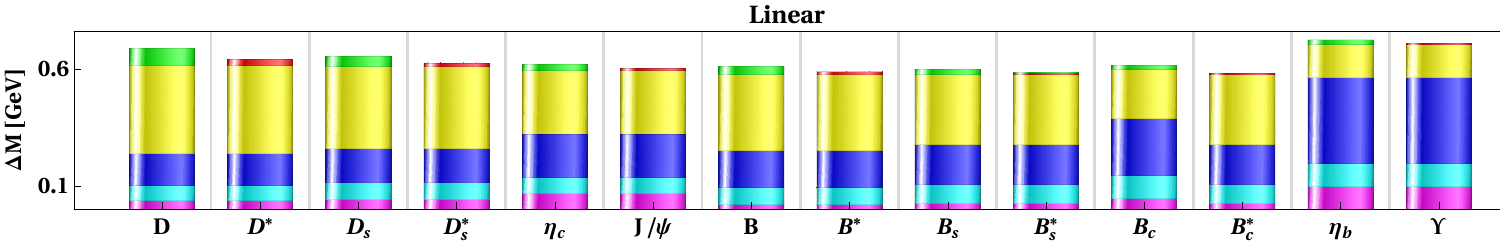}
\includegraphics[width=\linewidth]{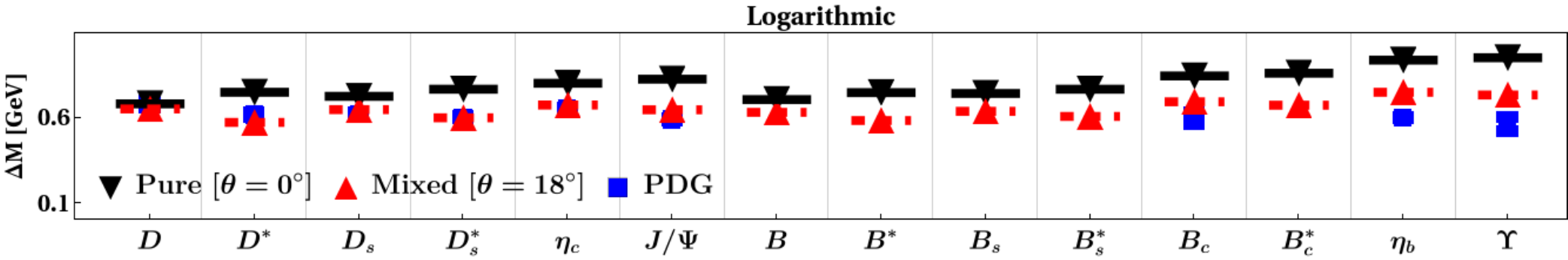}
\caption{The upper panel shows the various component contribution to the mass gap  $(\Delta M)$  from the present study with the logarithmic confinement whereas the middle panel presents the same with linear confinement from Ref. ~\cite{Arifi2022}. The Kin 1 and Kin 2 represents the heavier and light quark contribution to kinetic energy respectively. The red segment is considered as a subtracted part to the total mass gap since hyperfine vector contribution is negative. The lower panel shows the total mass gap from pure (solid-black) and mixed (red-dashed) configurations for $1S$ and $2S$ states where comparison is made with PDG quoted values (blue-squares) ~\cite{PDG2022}.} 
\label{Fig:contri}
\end{figure*}

\begin{figure*}
\includegraphics[width=\linewidth]{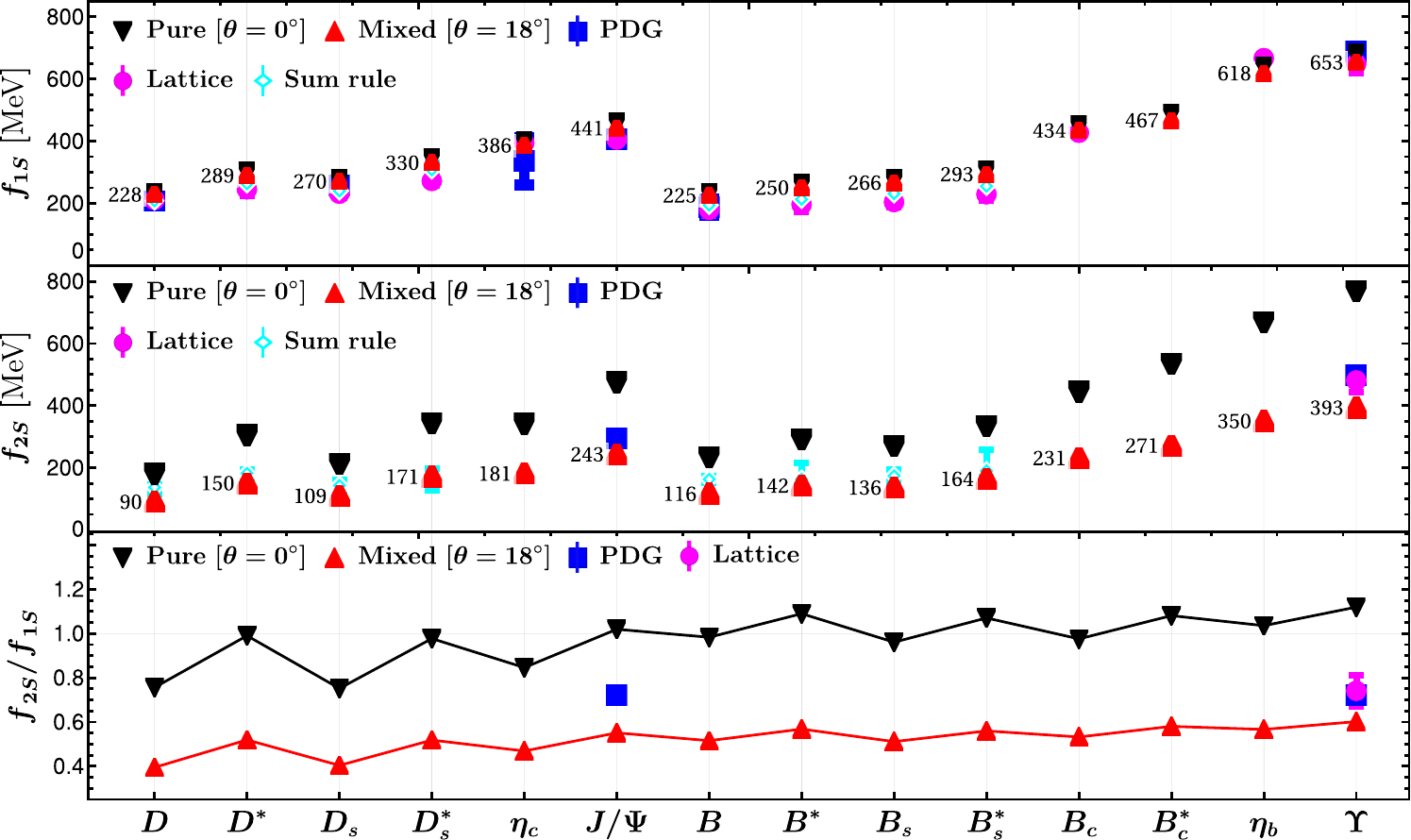}
\caption{The predicted decay constant of the $1S$ and $2S$ states of vector and pseudoscalar states. The results are compared with experimental values  listed by PDG~\cite{PDG2022}, available lattice simulations~\cite{Hatton2020,Hatton2021,Aoki2022} and SumRule prediction \cite{Bevcirevic2014,Veliev2011,Narison2002,Wang2015}.}
\label{Fig:Decay_Consts}
\end{figure*}

\begin{table*}
\begin{center}
\tabcolsep 2pt
\small
\begin{threeparttable}
\caption{Ground state mass spectra of mesons (All are in units of MeV).}
\label{sw2}
\begin{tabular}{lcccccccrcc}
\hline \hline
State  & Pure & Mixed $(18^{\circ})$ &  PDG \cite{PDG2022} & Linear \cite{Arifi2022} & Linear Mixed $(12^{\circ})$ \cite{Arifi2022} & Exponential \cite{Dhiman2019} & GI  \cite{Godfrey1985} & RQM \cite{Ebert:2002pp,Ebert:2009ua} \\
\hline
$\Upsilon(1S)$ &  9410 & 9390  & 9460.40 $\pm$ 0.10 & 9485 & 9480 & \ldots & 9460  &  9460\\
$\eta_b(1S)$ &  9381  & 9362  & 9398.7 $\pm$ 2.0 & {\textbf{9399}} & 9399 & \ldots & 9400 & 9400\\ 
$\Upsilon(2S)$ &  10365 & 10126 & 10023.4 $\pm$ 0.5 & 10377 & 10175 & \ldots & 10000 & 10023 \\
$\eta_b(2S)$ &  10322 & 10114 & 9999 $\pm$ 4 & 10249 & 10123 & \ldots & 9980 & 9993\\ 

$B^*_c(1S)$ & 6295  & 6285  & \ldots & 6343 & 6340 & \ldots & 6340 & 6332 \\
$B_c(1S)$ & 6262  & 6252  &  6274.47 $\pm$ 0.32 & 6269 & 6270 & \ldots & 6270 & 6270   \\ 

$B^*_c(2S)$ & 7159  & 6961 & \ldots & 7059 & 6930 & \ldots & 6890 & 6881 \\
$B_c(2S)$ & 7110 & 6948 & 6871.2 $\pm$ 1.0 & 6948 & 6885 & \ldots  & 6850 & 6835\\ 

$B^*_s(1S)$ & {\textbf{5416}} & {\textbf{5416}} & 5415.4$^{+1.8}_{-1.5}$ & 5421 & 5418 & 5329 & 5450 & 5414 \\
$B_s(1S)$ & {\textbf{5366}} & {\textbf{5366}} & 5366.92 $\pm$ 0.10 & 5325 & 5330 & 5313 & 5390 & 5392\\ 

$B^*_s(2S)$ & 6186  &  6026 & \ldots & 6067 & 5987 & \ldots & 6010 & 5992  \\
$B_s(2S)$ & 6111  & 6006 & \ldots & 5924 & 5928 & \ldots & 5980 & 5976 \\ 

$B^{*}(1S)$ & 5325 & 5327  & 5324.71 $\pm$ 0.21 & {\textbf{5325}} & 5325 & 5242 & 5370 & 5326 \\
$B(1S)$ & 5242 & 5247  & 5279.34 $\pm$ 0.12 & 5174 & 5182 & 5212 & 5310 & 5272 \\ 
$B^{*}(2S)$ & 6075 & 5913 & \ldots & 5968 & 5886 & \ldots & 5930 & 5906 \\
$B(2S)$ & 5951 & 5881 & {5869 $\pm$ 9}~$^{2}$ & 5740 & 5794 & \ldots & 5900 & 5890 \\ 

$J/{\Psi}(1S)$ & 3072 & 3065 & 3096.900 $\pm$ 0.006 & 3090 & 3087 & \ldots & 3100 & 3096\\
$\eta_c(1S)$ & 3023 & 3017  & 2983.9 $\pm$ 0.4 & 2987 & 2990 & \ldots & 2970 & 2979\\ 
${\Psi}(2S)$ & 3901   & 3714 & 3686.10 $\pm$ 0.06 & 3781 & 3670 & \ldots & 3680 & 3686\\
$\eta_c(2S)$ & 3828 & 3694 & 3637.7 $\pm$ 1.1 & 3627 & 3608 & \ldots & 3620 & 3588\\ 

$D^*_s(1S)$ &  2121 & 2121 & 2112.2 $\pm$ 0.4 & 2113 & 2111 & 1971 & 2130 & 2111 \\
$D_s(1S)$ & 2037  & 2040  & 1968.35 $\pm$ 0.07 & 1938 & 1946 & 1929 & 1980 & 1969\\ 
$D^*_s(2S)$ & 2891 & 2723  & 2714 $\pm$ 5 & 2798 & 2706 & \ldots & 2730 & 2731 \\
$D_s(2S)$ & 2765  & 2690 & {2591 $\pm$ 6}~$^{3}$ &  2546 & 2600 & \ldots & 2670 & 2688 \\ 

$D^{*}(1S)$ & 2022 & 2025 & $2010.26\pm05$ & 2020 & 2017 & 1884 & 2040 & 2010 \\
$D(1S)$ & 1884 & 1891 & $1869.66\pm05$ & 1731 & 1745 & 1803 & 1880 & 1871 \\
$D^{*}(2S)$ & 2774 & 2600 & $2627\pm 10$ & 2714 & 2608 & \ldots & 2640 & 2632 \\
$D(2S)$ & 2568 & 2545 & $2549\pm 19$ &  2282 & 2432 & \ldots & 2580 & 2581 \\ 

\hline \hline
\label{Tab:mass_spectra}
\end{tabular}
\begin{tablenotes}
\vspace{-0.3cm}
\footnotesize
\item[2] From the observation by LHC$_b$ Collaboration in $B$ mesons~\cite{LHCb2015}.
\item[3] From the recent observation by LHC$_b$ Collaboration in $D_s$ mesons~\cite{Lhcb2021ds}.
\end{tablenotes}
\end{threeparttable}
\end{center}
\end{table*}

\begin{table*}
\begin{center}
\tabcolsep 10pt
\small
\caption{Decay constant of heavy-heavy mesons (All are in units of MeV).} 
\label{decayconstanthh}
\begin{tabular}{lccccccc}
\hline 
\hline
 & $f_{\Upsilon}(1S)$ & $f_{\eta_b}(1S)$  & $f_{J/\psi}(1S)$  &   $f_{\eta_c}(1S)$ & $f_{B^*_c}(1S)$ & $f_{B_c}(1S)$ & References \\
\hline
Pure & 686 & 646 & 465 & 403 & 494 & 455 & This work\\
Mixed $\left(\theta=18^{\circ}\right)$ & 653 & 618 & 441 & 386 & 467 & 434 & This work\\
PDG  & 689 $\pm$ 5 & \ldots & 407 $\pm$ 5 & 335 $\pm$ 75 & \ldots & \ldots & \cite{PDG2022}\\
BLFQ & 532 & 574 & 379 & 423  & \ldots &\ldots & \cite{Li2016}\\
LFQM & 611 & 605 & 361 & 353 & 391 & 389 & \cite{Choi2015}  \\
LFQM (pure) & 688 & 647 & 403 & 356 & 436 & 406  & \cite{Arifi2022} \\
LFQM $(\theta=12^{\circ})$ & 666 & 629 & 390 & 347 & 421 & 393  & \cite{Arifi2022} \\
Lattice (HPQCD) & 677.2 $\pm$ 9.7 & 724 $\pm$ 24 & 410 $\pm$ 17 & \ldots & 422 $\pm$ 13 & 434 $\pm$ 15 & \cite{McNeile:2012qf,Donald:2012ga,Davies:2010ip,Colquhoun:2014ica,Becirevic:1998ua}\\ 
Lattice $n_f$ =2  & \ldots & \ldots & 418 $\pm$ 8 & 387 $\pm$ 7 & \ldots & \ldots & \cite{Bailas2018} \\ 
BS & 422  & 472  & 304  & 278 & 305 & 312 & \cite{Chen2019} \\
PM (Linear) & 706 & 744 & 338 & 363 & 435 & 465 & \cite{Patel2009} \\ 
QCDSR & \ldots & 251 $\pm$ 72 & 401 $\pm$ 46 & 309 $\pm$ 39 & \ldots & 320 $\pm$ 95 &\cite{Bevcirevic2014,Veliev2011}\\
\hline
 & $f_{\Upsilon}(2S)$ & $f_{\eta_b}(2S)$  & $f_{J/\psi}(2S)$  &   $f_{\eta_c}(2S)$ & $f_{B^*_c}(2S)$ & $f_{B_c}(2S)$ & References \\
 \hline
Pure & 768 & 669 & 474 & 341 & 534 & 444 & This work \\
Mixed $\left(\theta=18^{\circ}\right)$ & 393 & 350 & 243 & 181 & 271 & 231 & This work \\
Pure & 771 & 671 & 420 & 318 & 477 & 407 & \cite{Arifi2022} \\
Mixed $\left(\theta=12^{\circ}\right)$ & 498 & 443 & 274 & 214 & 308 & 268 & \cite{Arifi2022} \\
PDG & 497$\pm$5 & \ldots & 294 $\pm$ 5 & \ldots & \ldots & \ldots & \cite{PDG2022} \\
BLFQ & 518 $\pm$ 48 & 524 $\pm$ 58 & 312 $\pm$ 73 & 299 $\pm$ 68 & \ldots & \ldots & \cite{Li2017} \\
\hline 
\hline
\end{tabular}
\end{center}
\end{table*}

\begin{table*}
\begin{center}
\tabcolsep 3pt
\small
\caption{Decay constant of heavy-light mesons (All are in units of MeV).}
\label{decayconstanthl}
\begin{tabular}{lccccccccc}
\hline
\hline
  & $f_{B^*}(1S)$ & $f_{B}(1S)$  & $f_{B^*_s}(1S)$& $f_{B_s}(1S)$  & $f_{D^*}(1S)$ & $f_{D}(1S)$ & $f_{D^*_s}(1S)$  &  $f_{D_s}(1S)$ & References \\
\hline
Pure & 267 & 238 & 312 & 281 & 307 & 238 & 350 & 282 & This work\\
Mixed $(\theta = 18^{\circ} )$ & 250 & 225  & 293 & 266  & 289 & 228 & 330 & 270  & This work\\
BLFQ & 202 $\pm$ 3 & 233 $\pm$ 50 & 230 $\pm$ 36 & 259 $\pm$ 54 & 281 $\pm$ 20 & 295$\pm$ 63 & 306 $\pm$ 39 & 313 $\pm$ 67 &  \cite{Tang:2019gvn}\\
FLAG $N_f$ = 4 & \ldots & 190.0 $\pm$ 1.3 & \ldots & 230.3 $\pm$ 1.3 & \ldots & 212.0 $\pm$ 0.7 & \ldots & 249.9 $\pm$ 0.5 & \cite{Aoki2022}\\
QL (Taiwan) & \ldots & \ldots & \ldots & \ldots & \ldots & 235 $\pm$ 8 & \ldots & 266 $\pm$ 10 & \cite{Chiu2005} \\
LFQM (pure) & 215 & 196 & 256 & 235 & 265 & 212 & 303 & 251 &  \cite{Arifi2022} \\
LFQM $(\theta=12^{\circ})$ & 208 & 190 & 247 & 228 & 257 & 208 & 294 & 246 &  \cite{Arifi2022} \\
LFQM & 172 & 163 & 194 & 184 & 230 & 197 & 253 & 219 &  \cite{Dhiman2019}\\
LFQM  & 185 & 181 & 216 & 205 & 230 & 208 & 260 & 232 & \cite{Choi2015} \\ 
BS & 238 $\pm$ 18 & 196 $\pm$ 29 & 272 $\pm$ 20 & 216 $\pm$ 32 & 340 $\pm$ 23 & 230 $\pm$ 25 & 375 $\pm$ 24 & 248 $\pm$ 27 & \cite{cvetivc2004} \\
BS & \ldots & 193 & \ldots & 195 & \ldots & 238 & \ldots & 241 & \cite{wang2004} \\
QCDSR & \ldots & 203 $\pm$ 23 & \ldots & 236 $\pm$ 30 & \ldots & 203 $\pm$ 23 & \ldots & 235 $\pm$ 24 & \cite{Narison2002} \\
QCDSR & 213 $\pm$ 18 & 194 $\pm$ 15 & 255 $\pm$ 19 & 231 $\pm$ 16 & 263 $\pm$ 12 & 208 $\pm$ 10 & 308 $\pm$ 21 & 240 $\pm$ 10 & \cite{Wang2015}\\
RQM & 219 & 189 & 251 & 218 & 310 & 234 & 315 & 268 &  \cite{Ebert2006} \\
\hline
& $f_{B^*}(2S)$ & $f_{B}(2S)$  & $f_{B^*_s}(2S)$& $f_{B_s}(2S)$  & $f_{D^*}(2S)$ & $f_{D}(2S)$ & $f_{D^*_s}(2S)$  &  $f_{D_s}(2S)$ & References \\
\hline
Pure & 291 & 234 & 334 & 270 & 304 & 180 & 342 & 212 & This work \\
Mixed $(\theta = 18^{\circ})$ & 142 & 116 & 164  & 136 & 150 & 90  & 171 & 109 & This work \\
Pure & 236 & 197 & 276 & 232 & 266 & 168 & 301 & 200 & \cite{Arifi2022} \\
Mixed $(\theta = 12^{\circ})$ & 149 & 126 & 176  & 150 & 171 & 110  & 195 & 133 & \cite{Arifi2022}\\
RQM & \ldots & \ldots  & \ldots & \ldots & 293 & 292 & \ldots & \ldots & \cite{Shah2016bbs} \\
\hline
\hline
\end{tabular}
\end{center}
\end{table*}

\begin{table*}
\begin{center}
\tabcolsep 3pt
\small
\caption{The first six $\xi$ moments for the mixed $1S$ and $2S$ states of mesons.}

\label{ximoments}
\begin{tabular}{lcccccccccccccc}
\hline \hline
 & $\Upsilon(1S)$ & $\eta_b(1S)$ & $J/\psi(1S)$ & $\eta_c(1S)$ & $B_c^*(1S)$ & $B_c(1S)$ &  ${B^*}(1S)$ & ${B}(1S)$  & ${B^*_s}(1S)$ & ${B_s}(1S)$  & ${D^*}(1S)$ & ${D}(1S)$ & ${D^*_s}(1S)$  &  ${D_s}(1S)$  \\
\hline
$\langle \xi^1 \rangle$ & \ldots & \ldots & \ldots & \ldots & 0.378 & 0.378 & 0.607 & 0.604 & 0.580 & 0.579 & 0.311 & 0.301 & 0.270 & 0.267 \\
$\langle \xi^2 \rangle$ & 0.047 & 0.047 &  0.095 & 0.098 & 0.196 & 0.197  & 0.412 & 0.409 & 0.383 & 0.383 & 0.213 & 0.212 & 0.189 & 0.192\\
$\langle \xi^3 \rangle$ & \ldots & \ldots & \ldots  & \ldots & 0.111 & 0.112 &  0.299 & 0.297 & 0.272 & 0.272 & 0.132 & 0.132 & 0.106 & 0.108 \\
$\langle \xi^4 \rangle$ & 0.005 & 0.005 & 0.020 & 0.022 & 0.068 & 0.069 & 0.228 & 0.227 & 0.204 & 0.204 & 0.099 & 0.100 & 0.078 & 0.081 \\
$\langle \xi^5 \rangle$ & \ldots & \ldots & \ldots & \ldots & 0.044 & 0.044 & 0.180 & 0.180 & 0.158 & 0.159 & 0.073 & 0.075 & 0.054 & 0.057\\
$\langle \xi^6 \rangle$ & 0.001 & 0.001 & 0.006 & 0.006 & 0.029 & 0.030 & 0.147 & 0.146 & 0.125 & 0.127 & 0.059 & 0.061 & 0.042 & 0.044 \\
\noalign{\smallskip}\hline\noalign{\smallskip}
 & $\Upsilon(2S)$ & $\eta_b(2S)$ & $J/\psi(2S)$ & $\eta_c(2S)$ & $B_c^*(2S)$ & $B_c(2S)$ & ${B^*}(2S)$ & ${B}(2S)$  & ${B^*_s}(2S)$ & ${B_s}(2S)$  & ${D^*}(2S)$ & ${D}(2S)$ & ${D^*_s}(2S)$  &  ${D_s}(2S)$  \\
 \hline
 $\langle \xi^1 \rangle$ & \ldots & \ldots & \ldots & \ldots & 0.259 & 0.245 & 0.380 & 0.339 & 0.366 & 0.332 & 0.002 & -0.168 & 0.028 & -0.081 \\
$\langle \xi^2 \rangle$ & 0.093 & 0.098 & 0.180 & 0.208 & 0.154 & 0.151 & 0.158 & 0.114 & 0.165 & 0.131  & 0.073 & 0.010 &  0.132 & 0.119 \\
$\langle \xi^3 \rangle$ & \ldots & \ldots &  \ldots & \ldots & 0.097 & 0.098 & 0.064 & 0.024 & 0.083 & 0.055 & 0.009 & -0.047 & 0.051 & 0.034\\
$\langle \xi^4 \rangle$ & 0.014 & 0.015 & 0.050 & 0.060 & 0.069 & 0.071 & 0.020 & -0.014 & 0.047 & 0.025 & 0.020 & -0.013 & 0.060 & 0.061\\
$\langle \xi^5 \rangle$ & \ldots & \ldots & \ldots & \ldots & 0.050 & 0.052 & -0.001 & -0.030 & 0.030 & 0.012 & 0.010 & -0.016 & 0.041 & 0.042\\
$\langle \xi^6 \rangle$ & 0.003 & 0.003 & 0.017 & 0.021 & 0.037 & 0.040 & -0.011 & -0.035 & 0.021 & 0.008 & 0.012 & -0.005 & 0.039 & 0.044 \\
\hline \hline
\end{tabular}
\end{center}
\end{table*}
\begin{table*}
\begin{center}
\tabcolsep 6pt
\small
\caption{The charge radii $\langle r^2 \rangle$ of mixed 1S and 2S pseudoscalar mesons (All are in units of fm$^2$).}
\label{radius}
\begin{tabular}{lcccccccccc}
\hline \hline
& $\eta_b(1S)$  & $\eta_c(1S)$ &  $B_c^+(1S)$ &   ${B^0}(1S)$ &   ${B^+}(1S)$  & ${B_s^0}(1S)$ &  ${D^0}(1S)$ & ${D^+}(1S)$ &  ${D_s^+}(1S)$ & References \\
\hline
 This work & 0.009 & 0.039 & 0.032 & -0.125 & 0.246 & -0.067 & -0.231 & 0.144 & 0.085 &  \\
 LFQM & 0.010 & 0.042 & 0.036 & -0.155 & 0.314 & -0.080 & -0.282 & 0.171  & 0.095 & \cite{Arifi2022} \\
BLFQ  & 0.012 & 0.027 & \ldots & \ldots & \ldots & \ldots & \ldots & \ldots & \ldots & \cite{Li2017} \\
LFQM & \ldots & \ldots & 0.043 & -0.187 & 0.378 & -0.119 & -0.304 & 0.184 & 0.124 &  \cite{Hwang2002}   \\
CCQM & \ldots & \ldots & \ldots & \ldots & \ldots & \ldots & \ldots & 0.255  & 0.142 & \cite{Moita2021} \\
Lattice & \ldots & 0.063  & \ldots & \ldots & \ldots & \ldots & \ldots & \ldots  & \ldots & \cite{Dudek2006} \\
Lattice (L) & \ldots & \ldots  & \ldots & \ldots & \ldots & \ldots & \ldots & 0.138(13)  & \ldots & \cite{Can:2012tx} \\
Lattice (Q)  & \ldots & \ldots  & \ldots & \ldots & \ldots & \ldots & \ldots & 0.152(26)  & \ldots & \cite{Can:2012tx} \\
Lattice (B1)  & \ldots & 0.052(4)  & \ldots & \ldots & \ldots & \ldots & \ldots & 0.162(49) & 0.082(13) & \cite{Li:2017eic,Li:2020gau} \\
Lattice (C1)  & \ldots & 0.044(4)  & \ldots & \ldots & \ldots & \ldots & \ldots & 0.176(69) & 0.125(13) & \cite{Li:2017eic,Li:2020gau} \\
\hline
& $\eta_b(2S)$  & $\eta_c(2S)$ &  $B_c^+(2S)$ &   ${B^0}(2S)$ &   ${B^+}(2S)$  & ${B_s^0}(2S)$ &  ${D^0}(2S)$ & ${D^+}(2S)$ &  ${D_s^+}(2S)$ & \\
\hline
This work & 0.035 & 0.118 & 0.115 & -0.387 & 0.725 & -0.214 & -0.695 & 0.432 & 0.252 & \\
LFQM  & 0.030 & 0.129 & 0.118 & -0.450 & 0.911 & -0.222 & -0.801 & 0.464 & 0.266 & \cite{Arifi2022} \\
BLFQ  & 0.050 & 0.120 & \ldots & \ldots & \ldots & \ldots & \ldots & \ldots & \ldots & \cite{Li2017} \\
\hline \hline
\end{tabular}
\end{center}
\end{table*}

\section{Results and Discussion}\label{RD}
In the present study we have computed the ground state masses of heavy-heavy and heavy-light mesons within the LFQM employing the logarithmic confinement potential. The computed {spectroscopic results} are compared with the values listed by PDG~\cite{PDG2022} and previous LFQM model predictions based on linear confinement \cite{Arifi2022} and exponential confinement \cite{Dhiman2019}. {Additionally, comparisons are made with the GI model \cite{Godfrey1985} and the relativistic quark model (RQM) \cite{Ebert:2002pp, Ebert:2009ua}} in Table~\ref{Tab:mass_spectra}.  
We attempt to determine the variational parameters and quark masses that can systematically describe all mesons with different $q\bar{q}$ content, rather than the fine tunning of the spectra. As mentioned previously, we employ the experimental values for the $B_s^*(1S)$ and $B_s(1S)$ states provided by the PDG as input. The masses for other states are our model predictions. Notably, we observe that the calculated masses are found to be close to the experimental values of the respective states. 

In Fig.~\ref{Fig:mass spectroscopy1}, we show the mass spectra of the 1S and 2S {state} 
heavy mesons for both pure ($\theta=0^\circ$) and mixed ($\theta=18^\circ$) configurations. 
The black solid line with down-triangles represents the model predictions for logarithmic confinement 
without {the inclusion of} mixing effects, while the red dot-dashed line with up-triangles represents predictions after considering the mixing effect. The experimental values listed by the PDG~\cite{PDG2022} are shown with blue rectangles.
The green disks represent the input masses from the PDG used to fix the model parameters. {{Our findings clearly demonstrate that the spectroscopic results with mixing are more close to the existing experimental values listed in PDG \cite{PDG2022}. The impact of mixing appears {more} noticeable in the $2S$ states {compared} to $1S$ states.}}

{{The LHC$_b$ Collaboration has put forward both the natural $(0^-, 1^+, 2^-,...)$ and unnatural $(0^+, 1^-, 2^+,...)$ parity possibilities 
for {the} newly observed resonances $B_J(5840)$ and $B_J(5960)$~\cite{LHCb2015,PDG2022}. The average mass of 5869 $\pm$ 9 MeV for $B_J(5840)$ 
{excludes its identification as a}
higher orbital excited state ~\cite{LHCb2015,PDG2022}. Therefore, we {are} left with two possible assignments, $B(2S)$ and $B^*(2S)$, for $B_J(5840)$. 
Our prediction of $B(2S)$ as 5881 MeV {is} found to be comparable to {the} average measured mass of $B_J(5840)$. 
Hence, we identify $B_J(5840)$ as {the $B(2S)$ state of the} $B$ family. 
We note that {considering} $B_J(5960)$ as $B^*(2S)$ with {a} measured mass of 5969.2$\pm$ 2.9 MeV gives hyperfine splitting  $M(2^3S_1)-M(2^1S_0)$ {of} approximately 100 {MeV. This} is higher than {the well-established}  ground state hyperfine splitting $M(1^3S_1)-M(1^1S_0)$ of 45 MeV 
in $B$ mesons. This contradicts the typical expectation that the hyperfine splittings between the radially excited states should be lower 
than those {between the ground states}. Moreover, this consideration also violates the hierarchy  $\Delta M_{P} > \Delta M_{V}$.  Hence, from present study, we could identify $B_J(5840)$ as $B(2S)$. {However}, further experimental investigation and analysis may be necessary to reconcile {the} current understanding  
of {the} $2S$ states for heavy-light mesons. Also, the state $D_{s0}(2590)$ with {a mass of} 2591 $\pm$ 6  MeV was observed in $B^0 \rightarrow D^-D^+K^+\pi^-$ channel, decaying to $D^+K^+\pi^-$ final state by LHC$_b$~\cite{Lhcb2021ds}. {Given that the $K^+\pi^-$ system belongs to the $S$ wave below 
the $K^*(892)^0$ threshold, it indicates that} only the unnatural parity states decay to $D^+K^+\pi^-$~\cite{Lhcb2021ds}. These characteristics increases its identification as $D_s(2S)$. {Although our} predicted mass of 2690 MeV {differs by} roughly 100 MeV {from} the measured mass of $D_{s0}(2590)$, 
{it} still supports its identification as {the $D_s(2S)$ state of the} $D_s$ family. }}

It is also found that each confinement scheme displays a distinct level of predictability, as evidenced by the $\chi^2$ analysis. For the linear potential case, the $\chi^2$ analysis yields a value of 0.024 and 0.009 for pure and mixed cases, respectively~\cite{Arifi2022}. 
However, for the present study, we find the $\chi^2$ value of 0.019 for the pure scenario and 0.003 for the mixed one. 
Both studies suggests the necessity of mixing effects for achieving improved agreement with the experimental data. {The predictions from the logarithmic confinement scheme are
comparable to those from the linear confinement scheme somewhat even closer to the data in some cases, as illustrated in Table~\ref{Tab:mass_spectra}. 
By comparing LFQM spectroscopic predictions to those from the traditional potential models, {such as GI~\cite{Godfrey1985} and RQM~\cite{Ebert:2002pp,Ebert:2009ua}}, it is found that LFQM predictions are not {yet fully fine-tuned.} This is because the fixation of model parameters in general hadron spectroscopy refers to the process of determining the values of parameters{,} 
including constituent {quark masses}, coupling constants, potential parameters and other parameters depending on the specific system being considered{. These determinations span} from the light meson systems{,} including the pion known as the pseudo-Goldstone boson severely influenced by the chiral symmetry{, to the heavy meson systems,} including the open charm and bottom mesons described by the heavy quark symmetry.
One may separate the analysis {of} light meson sectors from the analysis {of} heavy meson sectors{,} 
as we focus {on the latter} in the present work. 
By adjusting the values of the model parameters to the known ground states of {a specific system, the spectra of radially and orbitally excited states are predicted}. Some of these models use   
the so called ``canonical" value of the strong coupling constant for heavy quarks, which is approximately $\alpha_s$ = 0.2, or other values in the close vicinity of this estimate~\cite{Lucha1991}. While the models following ``soft QCD" takes the value  $\alpha_s$ = 0.5 {to describe} light to heavy mesons in a unified manner~\cite{Godfrey1985}, 
more successful models {utilize} the strong ``running" coupling constant, taking into account how the strength varies with the changes in the energy scale. The discrepancy between our results and {the} PDG reported values of ground {state} masses may partly be attributed to the limitations of the present framework{, which does not include} fine-tuning of different energy {levels} specific to each system. As the constituent quarks inside the mesons have a mass difference {of the order of} hundred MeV, it may be too difficult to accommodate all the mesons with {the} same potential parameters. Instead of considering {the} same values of $a$ and $\alpha_s$ for all mesons, one may tune them 
{to be flavor- and scale-dependent} for further analysis beyond the present work.}}

{Figure~\ref{Fig:LFWFs}} illustrates the effects of mixing on the radial wave functions of 1S and 2S meson states. The upper panel compares the pure and mixed radial wave functions of the 1S and 2S states, as well as the ratios of the 2S to 1S states for both pure, $\phi^2_{2S}/\phi^2_{1S}$, and mixed, $\Phi^2_{2S}/\Phi^2_{1S}$, states in bottomonium ($b\bar{b}$). {Our results are also compared to those obtained from the linear confinement potential case~\cite{Arifi2022}.} 
{{The wave functions for $nS$ {states increase with an increase in} $n$ for both linear and logarithmic potential without mixing effects, implying that the decay constant would increase for larger $n$. This behaviour naively violates the requirement of the aforementioned experimental hierarchies. {However, the} inclusion of mixing significantly alters the radial wave functions of the 2S states, while the changes in the 1S states are minimal.}} 
The lower panel displays the mixed radial wave functions for heavy-heavy ($b\bar{b}, b\bar{c}$) and heavy-light ($b\bar{s}, b\bar{q}$) quark states {obtained from the logarithmic potential}. The range of the radial wave function is inversely proportional to the variational parameter $\beta$~\cite{Isgur:1988gb,Faiman:1968js}. From Table~\ref{Tab:variational_parameters}, it can be observed that bottomonium has the largest value of $\beta$, {resulting in} a narrower wave function compared to other meson states. Another noticeable observation is that the radial wave functions at the origin are proportional to {$\beta$, leading to} both the 1S and 2S states of bottomonium having the highest amplitudes at the origin among {the} meson wave functions. However, {the ratios of the 2S to 1S states remain constant,} indicating that {they are} independent of the quark flavor content.

{
For the analysis of the mass gap, defined as $\Delta M_{P(V)}= M^{2S}_{P(V)}-M^{1S}_{P(V)}$,  we decompose it as follows:
\begin{align}\label{Eq:Massgap_components}
\Delta M = \Delta M^{{\text{Kin}}} + \Delta M^{{\text{Conf}}}  + \Delta M^{{\text{Coul}}}  + \Delta M^{{\text{Hyp}}},
\end{align}
where the total mass gap is separated into four distinct contributions: kinetic energy ($\Delta M^{\rm Kin}$),  confinement potential ($\Delta M^{\rm Conf}$), 
Coulomb potential ($\Delta M^{\rm Coul})$ and hyperfine interaction ($\Delta M^{\rm Hyp}$). This decomposition facilitates a detailed analysis in our numerical calculations.
In Fig.~\ref{Fig:contri}, we present the decomposition of these contributions to the mass gap specifically for the mixed  states ($\Phi_{1S}, \Phi_{2S}$).} 
{The upper panel illustrates the contributions from the mass gap components for the logarithmic potential, while the middle panel shows an analogous taxonomic analysis for the linear potential~\cite{Arifi2022} for comparison.}
Here, 
{Kin 1 and Kin 2 represent the kinetic energy} contributions from the heavier quark and lighter quark to the kinetic energy, respectively. 
{Although} the mass gap $\Delta M$ appears to be almost flavor-independent, {its flavour dependency can be reasonably reflected through the} relative contributions from $\Delta M^{\text{Kin}}$, $\Delta M^{\text{Coul}}$ and $\Delta M^{\text{hyp}}$. The {Kin 1 and Kin 2} 
contribution is identical for quarkonia, whereas the {Kin 2} contribution increases {with} the mass difference between the constituent quarks 
{in mesons, as one might} intuitively anticipate.}
Moreover, as inferred from Eq.~\eqref{eq:mass_eigen_value}, it becomes apparent that $\Delta M^{\text{Coul}} \propto \beta$, indicating that the Coulomb contribution to the mass gap in heavy-light mesons is comparatively less than that {in} heavy-heavy mesons. This is consistent with the fact that at short distances, the dominant force is the attractive Coulomb interaction, {as illustrated by the blue regions in} the upper and middle panels of Fig.~\ref{Fig:contri}. {Indeed, {the contributions from the kinetic and Coulomb terms are minimal for} a logarithmic potential compared to a linear potential.}
The confinement contribution $\Delta M^{\text{Conf}}$ has no substantial dependency on $\beta$ (see Eq.\eqref{eq:mass_eigen_value}) for the logarithmic potential case. 
This {results in} an identical confinement contribution for heavy-light {and} heavy-heavy mesons, {making it} less comparable for different flavour mesons but more tractable{, as indicated by} the yellow regions in the upper panel of Fig.~\ref{Fig:contri}. 
It therefore reflects the flavour independence of the logarithmic potential. The confinement contribution to the mass gap associated with a linear potential varies inversely with $\beta$ ($\Delta M^{\text{Conf}} \propto \beta^{-1})$, capturing the dominant confinement forces at larger distances ~\cite{Arifi2022}.
The lower panel shows the mass gap $\Delta M$ between the 1S and 2S meson states. Here, the solid black line and red-dashed line represent our findings for pure and mixed configurations, respectively. The empirical mass constraint $\Delta M_{P} > \Delta M_{V}$ is found to be satisfied {only after} the inclusion of the mixing. 

{{In {the upper panel of} Fig.~\ref{Fig:Decay_Consts}, we present our findings for the decay constants {of} the 1S state with pure $(\theta = 0^{\circ})$ and mixed $(\theta = 18^{\circ})$ wave functions {,represented} by black down triangles and red up triangles, respectively. The similar results for $2S$ states are shown in the middle panel. In both the upper and middle panels, {the} numerical values {indicate} the decay constants obtained from the mixing of wave functions. For comparison we also show the available experimental data from PDG ~\cite{PDG2022}, other theoretical findings from Lattice ~\cite{McNeile:2012qf,Donald:2012ga,Davies:2010ip,Colquhoun:2014ica,Becirevic:1998ua} and QCDSR~\cite{Gelhausen:2014jea}. Apparently, the decay constants of the $1S$ states exhibit limited sensitivity {to} mixing effects, whereas those of the $2S$ states demonstrate significant changes. The $2S$ state decay constants, {when accounting for mixing effects are found to be closer} to other predictions~ \cite{PDG2022,McNeile:2012qf,Donald:2012ga,Davies:2010ip,Colquhoun:2014ica,Becirevic:1998ua,Gelhausen:2014jea}. 

To make a detailed comparison of the numerical results, we include our findings alongside other theoretical model calculations of decay constants for heavy-heavy~\cite{Li2016,Choi2015,McNeile:2012qf,Donald:2012ga,Davies:2010ip,Colquhoun:2014ica,Becirevic:1998ua,Arifi2022,Bailas2018,Chen2019,Patel2009,Bevcirevic2014,Veliev2011,Li2017} and heavy-light~\cite{Tang:2019gvn,Aoki2022,Arifi2022,Dhiman2019,Choi2015,cvetivc2004,wang2004,Narison2002,Wang2015,Ebert2006,Shah2016bbs} {mesons} in Tables~\ref{decayconstanthh} and~\ref{decayconstanthl}, respectively. From Table~\ref{decayconstanthh}, it is {evident} that our predicted {values for the} vector and pseudoscalar {meson} decay constants of quarkonia and $B_c$ mesons are {in} good agreement with the experimental results~\cite{PDG2022}. 
We also notice that{, for quarkonia, the condition $f_{1S} > f_{2S}$ is consistent with experimental observations when the mixing of wave functions is taken into account. Additionally,} it is seen that results from all LFQM with different confining schemes~\cite{Arifi2022,Dhiman2019} {show similar} quantitative results 
{to those of} the present study. However, predictions from BLFQ~\cite{Li2016} and Linear potential model~\cite{Patel2009} are roughly 100 MeV different {from} all other predictions. {A similar pattern is observed} for heavy-light mesons in Table~\ref{decayconstanthl}.

{To date, the} PDG has only listed $2S$ decay constants for vector quarkonium states. {Based on} the known experimental values of decay constants, the ratio $f_{2S}/f_{1S}$ must be less than unity. The lower panel presents the ratio $f_{2S}/f_{1S}$ for both pure and mixed scenarios. Except for $D$, $D_s$ and $\eta_c$, this empirical constraint {is} found to be violated in the pure case, { while it is satisfied {for} all heavy-light to heavy-heavy mesons in the mixed case}. {Despite this, our estimations of the ratio are} found to {somewhat differ from both experimental observations and lattice results. Further experimental measurements of $2S$ state decay constants will be significant to better understand and observe the impact of mixing effects.
}}

\begin{figure*}[ht]
\centering
\includegraphics[width=\linewidth]{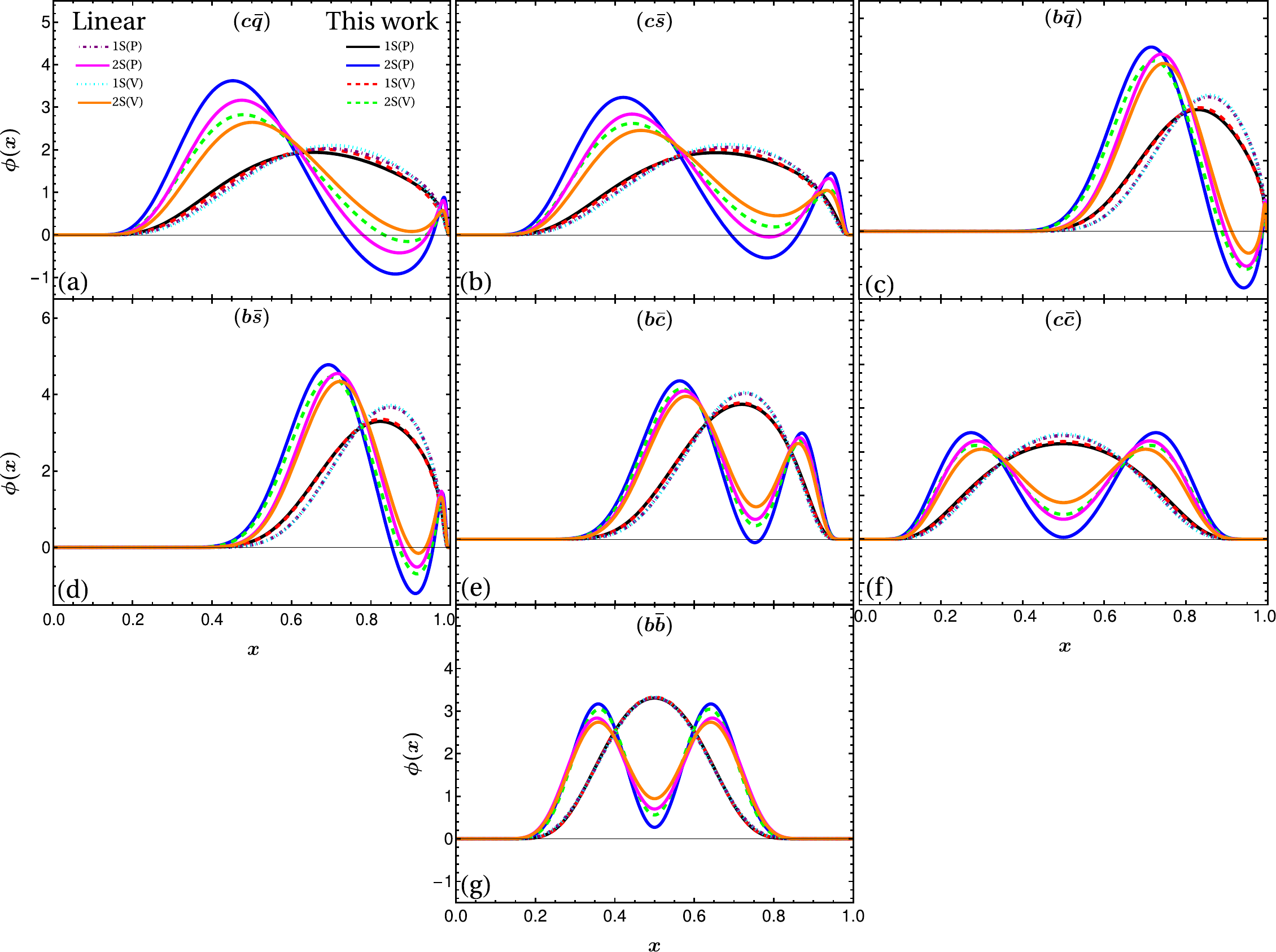}
 \caption{Twist-2 Distribution amplitude of $1S$ and $2S$ pseudoscalar and vector states for the case of mixed configuration with $\theta = 18^{\circ}$. The comparison with linear confining potential corresponds to $\theta = 12^{\circ}$ also presented~\cite{Arifi2022}. }
\label{DAFIG}
\end{figure*}

\begin{figure*}[ht]
\centering
\includegraphics[width=\linewidth]{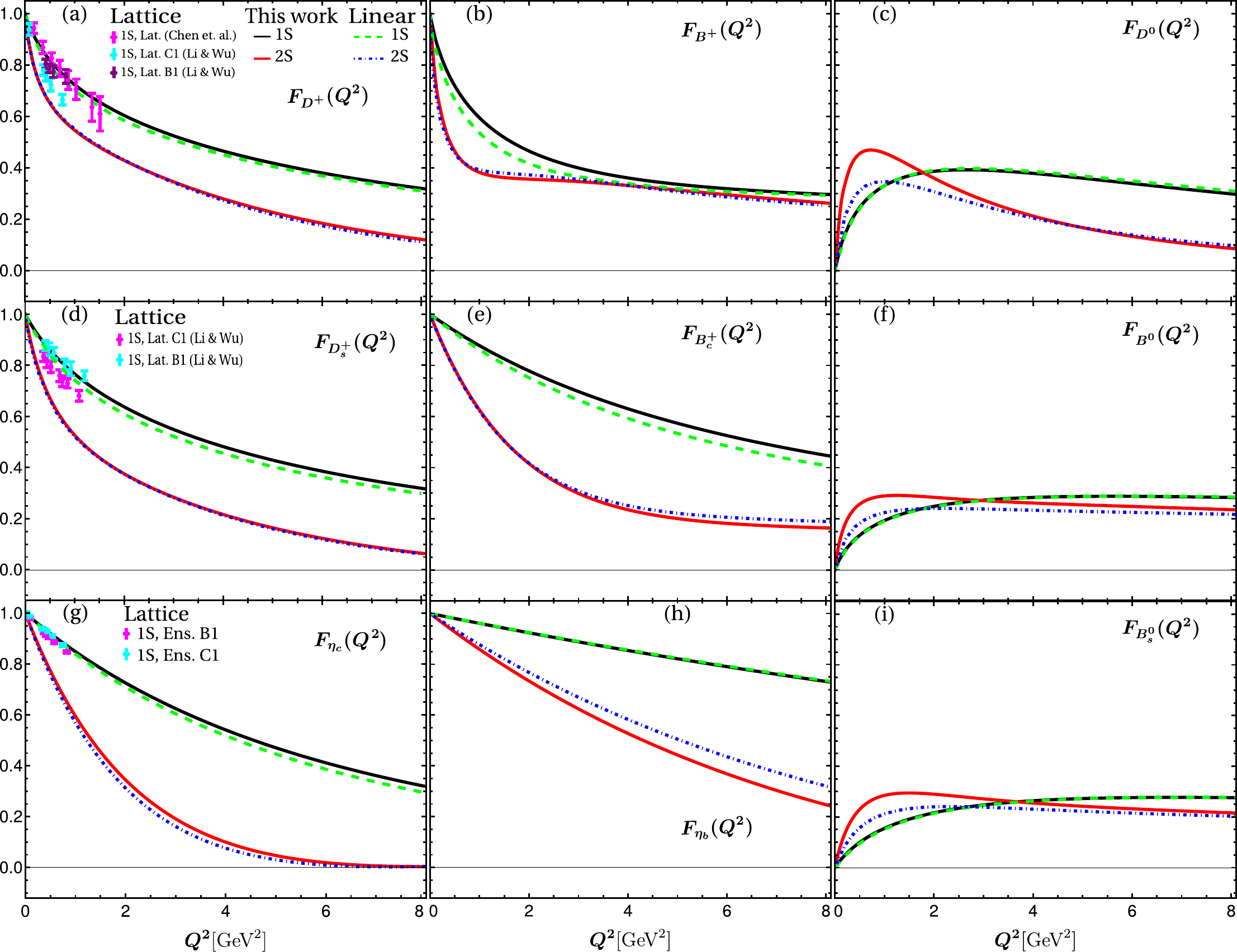}
\caption{Electromagnetic form factors of the $1S$ and $2S$ pseudoscalar states of mesons with the mixing angle $\theta = 18^{\circ}$. The linear confinement potential results with mixing angle $\theta = 12^{\circ}$~\cite{Arifi2022} and available lattice findings ~\cite{Can:2012tx,Li:2017eic,Li:2020gau} are also presented for comparison. For the case of quarkonium only the contribution from quark is considered.}
\label{EFFFIG}
\end{figure*}
{Figure}~\ref{DAFIG} shows the normalized twist-2 quark 
DAs for vector and pseudoscalar mesons with {a mixing angle of} $\theta = 18^{\circ}$. The solid black and solid blue lines correspond to {the} $1S$ and $2S$ pseudoscalar mesons, respectively, while the red dashed and green dashed lines represent {the} $1S$ and $2S$ vector meson states. The DAs without and with {the} inclusion of mixing effects in {the} wave function {do not show} significant differences, so we do not include results with $\theta = 0^{\circ}$. For comparison, we also include the results from {the} linear confining potential~\cite{Arifi2022}. 
We follow the traditional convention where the heavier quark inside the mesonic system carries the longitudinal momentum fraction $x$ while the the lighter quark carries the fraction $1-x$.
The DAs for the $1S$ pseudoscalar and vector states are not significantly different {because the} variational parameter $\beta$ is {the} same for both states within the present formalism. It is noted that the DAs {for} bottomonium and charmonium {exhibit} symmetry under the change {from} $x$ to $1-x$. However, {the} charmonium DA is found to be wider {compared to bottomonium DA due to its lower mass,} as expected. Similarly, the {extrema} for charmonium {are} shifted more towards the endpoints {compared} to bottomonium. {In} the case of heavy-light mesons, as the mass difference between the constituent quark and anti quark inside meson increases, their DAs appear more asymmetrical and sharply peaked. The impact of mass on the DAs can also be noticed from the heavy-light systems. In particular, bottom and bottom-strange mesons have narrower DAs than the charm and charm-strange mesons.
We observed similar behavior in the twist-2 DAs compared to those obtained using a linear confining potential. Our predictions show both qualitative and quantitative agreement with these results. It is also apparent from Fig.~\ref{DAFIG} that the $2S$ pseudoscalar and vector DAs are {distinct in contrast to the $1S$ states}. This trend {is} found to be contradictory to {the BLFQ results, which indicate that the $1S$ DAs tend} to be more distinct~\cite{Tang:2019gvn}. We also find significant difference in the {minima of the present DAs compared to the} BLFQ outcomes~\cite{Tang:2019gvn}.

Table \ref{ximoments} shows the $\xi$ moments up to $n$ = 6 for vector and pseudoscalar states of $1S$ and $2S$. {Due to the same} constituent quark masses and symmetrical DAs, the odd-$n$ moments for quarkonium {are} found to be zero. {In} the case of heavy-light mesons, due to mass difference between constituent quarks odd-$n$ moments also exhibits the asymmetry.
For instance, $\langle \xi^1 \rangle_{B_c(1S)}$,  $\langle \xi^1 \rangle_{B_s(1S)}$ and $\langle \xi^1 \rangle_{B_q(1S)}$ are 0.378, 0.579 and 0.604{, respectively increasing} as the difference between constituents mass increases. The fact {that} 
$\langle \xi^1 \rangle$ of the $D(2S)$ and $D_s(2S)$ {states is} observed to be negative {is due to the} wide spread of their DAs in the 0 $<$ $x$ $<$ 0.5 domain. Overall, the predictions of $\xi$ moments are in reasonable agreement with the results {obtained from the} linear potential~\cite{Arifi2022}. 

In Fig.~\ref{EFFFIG}, we present the EMFFs of the $1S$ (black) and $2S$ (red) states of heavy-heavy and heavy-light pseudoscalar mesons, respectively. We also illustrate the lattice simulation results for comparison~\cite{Can:2012tx,Li:2017eic,Li:2020gau}. {For quarkonium}, we consider only the contribution from the quarks {due} to the similar flavour content. It is {noted} that our {predictions for the $1S$ states of} $D^{+},~D^{+}_{s}$ and $\eta_{c}$ mesons  are in good agreement with the lattice results. The {form factors of the $2S$ states show a steeper decline} compared to the $1S$ states. However, overall one sees a qualitative similarity in the behaviour of the FFs compared to a similar model with linear confinement~\cite{Arifi2022}. We observe that {for
heavy-light systems such as} $D^+$, $D_s^+$ and $B^+$, the dominant contribution to the form factor {in the} high $Q^2$ region comes from the heavy quark whereas the light quark contribution is negligible. Conversely, for the $B_c$ meson, both $b$ and $c$ quark contribute equally to the form factors. 
Table \ref{radius} shows the obtained charge radii $\langle r_{nS}^2 \rangle$ of the pseudoscalar {states} of heavy-heavy and heavy-light mesons{, along with other} available predictions~\cite{Can:2012tx,Li:2017eic,Li:2020gau}. We find that our results are in {relatively} good agreement {with other} model results~\cite{Li2017,Arifi2022,Hwang2002,Moita2021}.

In Fig.~\ref{TFFFIG1}, we present the TFFs for charmonium in the spacelike ($Q^2 < 0$) momentum transfer region. The 
left plot displays the 
asymptotic behavior for high $Q^2$ values.  
The comparison {is} made with the experimental data from $BABAR$~\cite{BaBar:2010siw} along with outcomes from a linear confining potential \cite{Arifi2022}. The 
right panel presents similar results for $2S$ state. {{The $1S$ {state TFFs exhibit} similar behaviour for both logarithmic and linear {potentials and appear}  insensitive to mixing effects. 
The {$Q^2 F_{\eta_c \gamma}(Q^2)$ demonstrates} excellent agreement with the experimental data{, especially in the} region of $Q^2 < 15~ \text{GeV}^2$ for $1S$ state~\cite{BaBar:2010siw}. It is noted that the mixing effects are apparent in $Q^2 F_{\eta_c \gamma}(Q^2)$ {for the} $2S$ state.}} 

{Figure~\ref{TFFFIG2} shows} the 
TFFs for bottomonium in the spacelike momentum transfer region. The line codes are the same as in Fig.~\ref{TFFFIG1}. 
Although the qualitative behavior of $F_{\eta_{b}\gamma}$ resembles that of $F_{\eta_{c}\gamma}$, their quantitative characteristics, such as the slope of the form factor at $Q^{2}=0$, differ significantly due to the heavier $b$ quark compared to the $c$ quark. {{Again, the mixing effects can be seen here for {the} $2S$ state in {$Q^2 F_{\eta_b \gamma}(Q^2)$}. However, the other quantities remain unaffected for the choice of potential and mixing of wave functions. Further experimental measurements can provide more insights and refine our current understanding of TFFs.}}

The di-leptonic decay widths of the charged pseudoscalar mesons are presented in Table \ref{leptonict}. {{The uncertainty in the prediction of these decay widths from the present study and the linear potential model is due to the uncertainty in the values of CKM matrix elements.}} The estimated widths for these transitions are approximately twice as large as those reported by~\cite{ciftci2000}.  Also, we note that the large uncertainty in the experimental BRs for some transitions leads to inconclusiveness in the results. 
However, 
{
our predictions for
 ${\rm Br}(B^+ \rightarrow e^+ {\nu_e})=(1.30 \pm 0.14) \times 10^{-11}$, ${\rm Br}(B^+ \rightarrow \mu^+ {\nu_{\mu}})=(5.75 \pm 0.60)\times 10^{-7}$, ${\rm Br}(D^+ \rightarrow e^+ {\nu_e})=(1.02 \pm 0.04) \times 10^{-8}$ and ${\rm Br}(D_{s}^+ \rightarrow e^+ {\nu_e})=(1.45 \pm 0.02) \times 10^{-7}$ 
}are in accordance with the experimental BR limits $(<9.8 \times 10^{-7})$,  $(<8.6 \times 10^{-7})$,  $(<8.8 \times 10^{-6})$ and  $(<8.3\times 10^{-5})$, respectively, as quoted by the PDG \cite{PDG2022}. 
{
We also find that our predictions for ${\rm Br}(B^+ \rightarrow \tau^+ {\nu_{\tau}})=(1.29 \pm 0.13) \times 10^{-4}$ and ${\rm Br}(D^+ \rightarrow \tau^+ {\nu_{\tau}})=(1.76 \pm 0.06) \times 10^{-3}$
}
are close to the experimental values of $(1.09 \pm 0.24) \times 10^{-4}$ and $(1.20 \pm 0.27) \times 10^{-3}$, respectively. 

The rare decays of the charge neutral bottom and bottom-strange mesons by considering the uncertainty persist into the CKM matrix elements are shown in Table \ref{rare}.  Our results for $B^0 \rightarrow l^{+}l^{-} (l=e,\mu, \tau)$ are consistent with the results of \cite{bobeth2014}. The search and analysis of rare decays are challenging experimentally due to the considerably small branching fractions. The large uncertainty associated with the experimental BRs for these processes make it difficult to draw reliable conclusion. However, 
{our predictions for ${\rm Br}(B^0\rightarrow e^{+}e^{-})=(3.93 \pm 0.41 ) \times 10^{-15}$ and ${\rm Br}(B_s^0\rightarrow e^{+}e^{-})=(1.32 \pm 0.13) \times 10^{-13}$ fall}
within the limits of $ < 3.0  \times 10^{-9}$ and $ < 11.2 \times 10^{-9}$, respectively, as measured very recently by {\small{LHC$_b$}}~\cite{Aaij2020}. 
Also, { our prediction of ${\rm Br}(B^0\rightarrow \mu^{+}\mu^{-})=(1.73 \pm 0.18) \times 10^{-10}$} falls within the  range of the values listed by LHC$_b$ $(< 2.6 \times 10^{-10})$ \cite{Aaij2022, Aaij20221} and CMS $(<3.6 \times 10^{-10})$ \cite{sirunyan2020} Collaboration. Similarly, { our prediction of ${\rm Br}(B_s^0 \rightarrow \mu^{+} \mu^{-})=(5.81 \pm 0.58) \times 10^{-9}$} is in excellent agreement with the experimental observation of $(3.09^{+0.46}_{-0.43}) \times 10^{-9}$ \cite{Aaij2022, Aaij20221} reported by LHC$_b$ and $(2.9 \pm 0.7 \pm 0.2 )\times 10^{-10}$ by CMS \cite{sirunyan2020}. 
{{ {We also obtain the ratio $\frac{{\rm Br}(B^0 \rightarrow \mu^{+} \mu^{-})}{{\rm Br}(B_s^0 \rightarrow \mu^{+} \mu^{-})}= 0.0298 \pm 0.0042$,} which is in excellent agreement with the} Standard Model prediction of $0.0295^{+0.0028}_{-0.0025}$~\cite{bobeth2014}. However, the combined analysis of CMS and LHC$_b$ quotes this ratio as $0.14^{+0.08}_{-0.06}$ that deviates by 2.3$\sigma$ from the standard model prediction \cite{CMS:2014xfa,bobeth2014}.} The estimated BR for $B^0 \rightarrow \tau^{+}\tau^{-}$ and $B_s^0 \rightarrow \tau^{+}\tau^{-}$ are in good agreement with those predicted by \cite{bobeth2014} and the experimental data from PDG \cite{PDG2022}. {{We note that weak decays from LFQM with different confinement schemes have predictions close to the measured values of the respective decay channels. However, more detailed analysis on this can be made by evaluating other observables like weak form factors, forward-bakward asymmetry, iso-spin asymmetry and longitudinal and transverse polarization fractions with high $q^2$ within LFQM.
}}

\begin{figure*}[ht]
\centering
\includegraphics[width=0.45\linewidth]{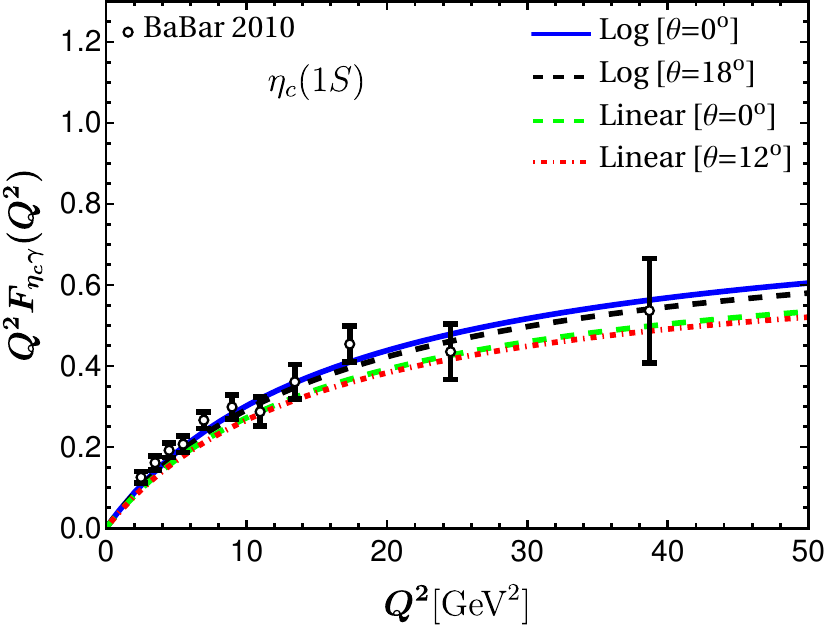}
\hspace{0.3cm} 
\includegraphics[width=0.45\linewidth]{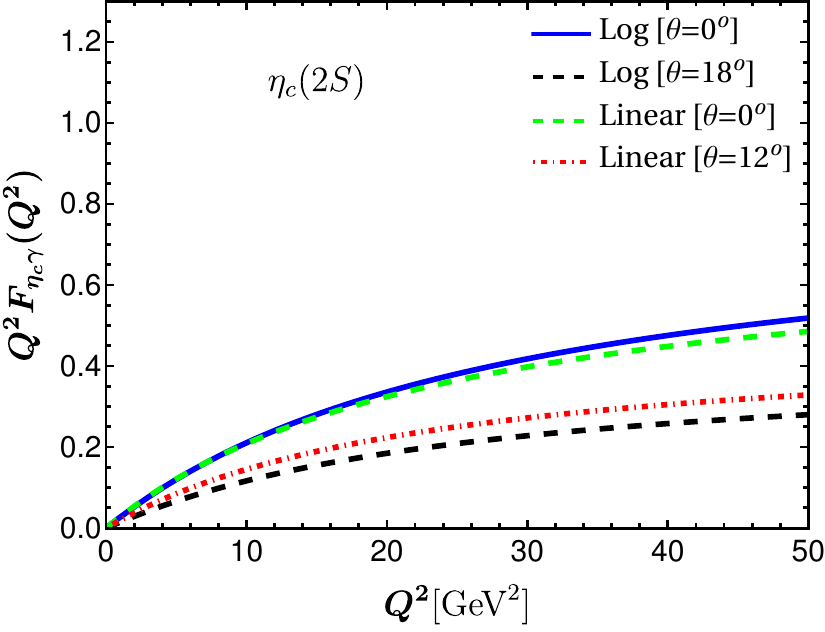}
\caption{Two photon transition form factors of pseudoscalar states $1S$ and $2S$ of charmonium from logarithmic potential with pure $(\theta = 0^{\circ})$ and mixed $(\theta = 18^{\circ})$ wave functions. The results obtained from linear confining potential parameters are also presented for pure $(\theta = 0^{\circ})$ and mixed $(\theta = 12^{\circ})$ cases~\cite{Arifi2022}. The experimental measurements are from $BABAR$ Collaboration~\cite{BaBar:2010siw}.}   
\label{TFFFIG1}
\end{figure*}
\begin{figure*}[ht]
\centering
\includegraphics[width=0.45\linewidth]{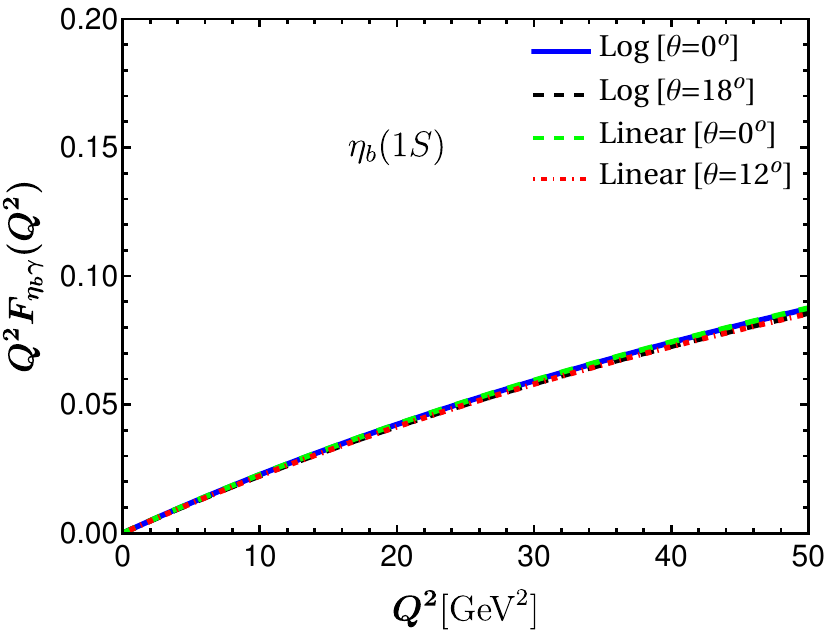}
\hspace{0.3cm}
\includegraphics[width=0.45\linewidth]{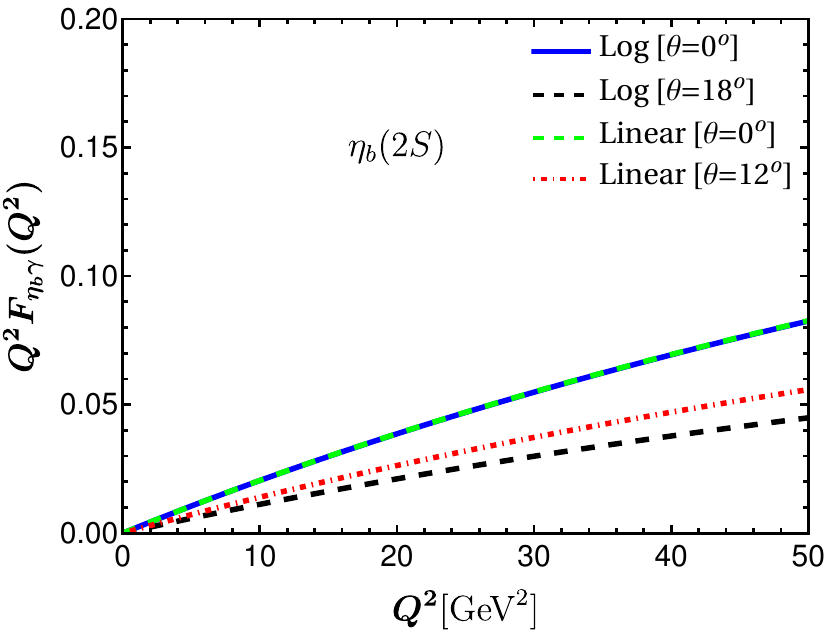}
\caption{Two photon transition form factors of pseudoscalar states $1S$ and $2S$ of bottomonium from logarithmic potential with pure $(\theta = 0^{\circ})$ and mixed $(\theta = 18^{\circ})$ wave functions. The results obtained from linear confining potential parameters are also presented for pure $(\theta = 0^{\circ})$ and mixed $(\theta = 12^{\circ})$ cases~\cite{Arifi2022}.} 
\label{TFFFIG2}
\end{figure*}

\begin{table*}
\begin{center}
\tabcolsep 16pt
\small
\caption{The branching ratios (in \%) for di-leptonic decays of heavy-light mesons. Experimental results are from PDG \cite{PDG2022}.}
\label{leptonict}
\begin{tabular}{lcccccc}
\hline \hline
Transition & This Work &   Linear \cite{Arifi2022} &   \cite{ciftci2000}  & Experiment \cite{PDG2022} \\
 &    $(\theta = 18^{\circ})$ &      $(\theta = 12^{\circ})$  &      &    \\
\hline
$B^+ \rightarrow e^+ {\nu_e}$  & $(1.30 \pm 0.14)\times 10^{-11}$ &  $(0.91 \pm 0.10 ) \times 10^{-11}$  &  $\cdots$  & $<9.8 \times 10^{-7}$ \\

$B^+ \rightarrow \mu^+ {\nu_\mu}$   & $(5.75 \pm 0.60) \times 10^{-7}$ & $(4.05 \pm 0.42) \times 10^{-7}$ &   4.82 $\times
10^{-7}$         &  $<8.6 \times 10^{-7}$ \\

$B^+ \rightarrow \tau^+ {\nu_\tau}$  &   $ (1.29 \pm 0.13) \times 10^{-4}$   & $(0.90 \pm 0.09) \times 10^{-4}$ &     9.25 $\times 10^{-5}$ &    $(1.09 \pm 0.24) \times 10^{-4}$ \\ 

$D^+ \rightarrow e^+ {\nu_e}$  & $ (1.02 \pm 0.04) \times 10^{-8}$ & $(0.78 \pm 0.03 )\times 10^{-8}$ &  \ldots  & $<8.8 \times 10^{-6}$ \\

$D^+ \rightarrow \mu^+ {\nu_\mu}$  &  $(4.47 \pm 0.16) \times 10^{-4}$   & $( 3.42 \pm 0.12) \times 10^{-4}$ &   2.87 $\times 10^{-4}$  &  $(3.74 \pm 0.17) \times 10^{-4}$   \\

$D^+ \rightarrow \tau^+ {\nu_\tau}$ & $(1.76 \pm 0.06) \times 10^{-3}$  & $(0.13 \pm 0.01)\times 10^{-3}$ &   0.75 $\times 10^{-3}$ &    $(1.20 \pm 0.27) \times 10^{-3}$ \\  

$D_s^+ \rightarrow e^+ {\nu_e}$  &  $(1.45 \pm 0.02) \times 10^{-7}$    & $(1.15 \pm 0.01) \times 10^{-7}$  &\ldots  & $<8.3 \times 10^{-5}$\\

$D_s^+ \rightarrow \mu^+ {\nu_\mu}$    &  $(6.39 \pm 0.08) \times 10^{-3}$   & $(5.05 \pm 0.06) \times 10^{-3}$ &   4.41 $\times 10^{-3}$       &  $(5.35 \pm 0.12) \times 10^{-3}$   \\

$D_s^+ \rightarrow \tau^+ {\nu_\tau}$  & $(10.71 \pm 0.13) \times 10^{-2}$ & $(4.02 \pm 0.05) \times 10^{-2}$  & 4.30 $\times 10^{-2}$ & $(5.36 \pm 0.10) \times 10^{-2}$ \\
\hline \hline
\end{tabular}
\end{center}
\end{table*}

\begin{table*}
\begin{center}
\tabcolsep 5pt
\small
\caption{The branching ratios (in \%) for rare di-leptonic decays of charge neutral heavy-light mesons. Experimental results are from PDG\cite{PDG2022}.}
\label{rare}
\begin{tabular}{lcccr}
\hline \hline
Transition &   This Work  &        Linear \cite{Arifi2022}  &    \cite{bobeth2014}  &   Experiment \cite{PDG2022}  \\
 &    $(\theta = 18^{\circ})$ &      $(\theta = 12^{\circ})$  &      &    \\
\hline
$B^0 \rightarrow e^+ e^-$  &   $(3.93 \pm 0.41 ) \times 10^{-15}$ &    $( 2.76 \pm 0.29 )  \times 10^{-15}$ &    $(2.48 \pm 0.21) \times 10^{-15} $ & $< 2.5 \times 10^{-9}$ \\
   &  &  &  & $< 3.0 \times 10^{-9}$ \cite{Aaij2020} $[LHCb]$ \\

$B^0 \rightarrow \mu^+ \mu^-$  &   $( 1.73 \pm 0.18) \times 10^{-10}$ &      $( 1.22 \pm 0.13 ) \times 10^{-10}$ &  $(1.06 \pm 0.09)\times10^{-10}$ &  $< 5^{+17}_{-15}  \times 10^{-11}$ \\

  &  &  &  & $< 2.6 \times 10^{-10}$ \cite{Aaij2022,Aaij20221} $[LHCb]$ \\ 

&  &  &  & $< 3.6 \times 10^{-10}$ \cite{sirunyan2020}$[CMS]$\\

$B^0 \rightarrow \tau^+ \tau^-$    & $(3.65 \pm 0.38 )\times 10^{-8}$    &    $(2.53 \pm 0.26) \times 10^{-8}$  &   $(2.22 \pm 0.19)\times10^{-8}$  &   $<2.1 \times 10^{-3}$ \\

$B_s^0 \rightarrow  e^+ e^-$    
& $(13.19 \pm 1.32 ) \times 10^{-14}$     &     $( 9.62  \pm 0.97 ) \times 10^{-14}$ &      $(8.54 \pm 0.55)\times 10^{-14}$ & $<9.4 \times 10^{-9}$ \\
   &  &  &  & $ < 11.2 \times 10^{-9}$ \cite{Aaij2020}$[LHCb]$\\  
$B_s^0 \rightarrow \mu^+ \mu^-$     & $(5.81  \pm 0.58 )\times 10^{-9}$  &     $( 4.24 \pm 0.43 ) \times 10^{-9}$ &       $(3.65 \pm 0.23)\times 10^{-9} $ &     $(2.9 \pm 0.4) \times 10^{-9}$ \\
 &  &  &  & $(3.09^{+0.46}_{-0.43} \pm 0.14) \times 10^{-9}$\cite{Aaij2022,Aaij20221}$[LHCb]$\\ 
   &  &  &  & $(2.9 \pm 0.7 \pm 0.2) \times 10^{-10}$\cite{sirunyan2020}$[CMS]$ \\ 
$B_s^0 \rightarrow \tau^+ \tau^-$ &   $(12.47 \pm 1.25 ) \times 10^{-7}$ &      $(9.05 \pm 0.91) \times 10^{-7}$ &  $(7.73 \pm 0.49) \times 10^{-7}$   & $<6.8 \times 10^{-3}$  \\
\hline \hline
\end{tabular}
\end{center}
\end{table*}

\section{summary}\label{SUM}
In this study, we examined the 1S and 2S states of heavy mesons using both pure and mixed harmonic oscillator wave functions. 
{A key aspect of our work is the adoption of the logarithmic plus Coulomb potential, which sets our {present work apart from the previous works with other potential models}.}
To differentiate between vector and pseudoscalar mesons, we exploited the variational principle 
{with this unique {logarithmic} potential, treating the hyperfine interaction term perturbatively}. 

Following the variational principle, we also derived a constraint for the model parameters.
Given {the} fixed quark masses, these model parameters were determined 
{ using two meson masses as inputs, allowing us} to predict and subsequently verify other observables like decay constants, twist-2 distribution amplitudes, electromagnetic form factors, {and} charge radii of the ground and radial excited states. We have also predicted the di-leptonic and rare decays of the pseudoscalar ground states.  

Our estimates for mass spectra are in good agreement with the existing experimental data listed by PDG.  While there is no discernible difference in the masses of 1S states of heavy mesons in the pure and mixed scenarios, the predicted masses of 2S states undergo notable modifications and demonstrate improved consistency with experimental values after introducing mixing. Moreover, based on the predicted masses, we assign $2^1S_0$ label to newly observed state $B_J(5840)$. 

It is noticed that the mass gap between the 1S and 2S states of mesons is roughly around 600 MeV, irrespective of the flavor content. 
{Within} the LFQM formalism, the mass disparities of pseudoscalar mesons can exceed those of vector mesons, independent of the quark flavor content, by employing a mixing angle greater than the critical angle $\theta = 6^{\circ}$. This can be identified by the relation presented in Eq. (\ref{eq:mass_gap}), which demonstrates that the hierarchy manifests in the opposite direction in the absence of mixing. For the case of decay constants, 
{the predictions of $1S$ states,} even without mixing effects, are found to be in good agreement with the experimental data. However, obtaining the correct sequence of decay constants for {the} 2S states necessitates accounting for mixing effects. 
This also emphasizes that mixing effects are crucial {to understanding} the characteristics of the $2S$ states.  

{It is worth noting that there are no disparities between the DAs of the pseudoscalar and vector states for the 1S states, whereas such disparities become more pronounced in the context of the 2S states.}
Furthermore, we observe {that the} DAs spread up to {a} few GeV for transverse momentum, with lighter quark-containing mesons displaying shorter tails. For {a} comprehensive analysis, we have computed the corresponding $\xi$ moments up to $n = 6$ in this study. Additionally, the charge radii and electromagnetic form factors for mesons are calculated and found to be similar to the available lattice simulation data. 
Since the wave functions of the $2S$ states are more widely distributed in space {due to} the mixing effects, the wave functions near the origin become smaller. The behavior of the decay constants, which are lowered by the mixing, is the opposite of this. The ${Q^2} F_{\eta_c(b) \gamma}(Q^2)$ for $2S$ states {shows} notable changes after {accounting for mixing effects,} whereas the $1S$ states remain unaffected. The BRs for various di-leptonic decays are found to be consistent with the existing experimental data. 

{The} present attempt to study LFQM with the logarithmic confinement for heavy-light to heavy-heavy mesons has been reasonably effective  {in demonstrating} a unified picture of meson spectroscopy when mixing effects are taken into account. 
Perhaps {the more sophisticated analysis utilizing the AI/ML technology may deserve further investigation to yield the more fine-tuned mass spectra and structure studies in the near future.}  Since the mixing effects of wave functions play a major role {as discussed in {the present} work, the future progress in the meson spectroscopy may be forthcoming in the direction of handling the mixing effects with the more realistic trial wave functions beyond the simple harmonic oscillator wave functions. The analysis of various structural properties associated with the meson wave functions deserves further study to understand the interplay of various states}. As a possible extension of present work, it would be encouraging to obtain the higher radial excited states {including 3S states} by expanding the wave function in terms of a larger basis to identify many of the newly observed resonances.  
 
\begin{acknowledgements}
B. P. expresses gratitude to P C Vinodkumar for valuable discussion on this work.
{
The work of H.-M.C. was supported by the National 
Research Foundation of Korea (NRF) under Grant No. NRF- 2023R1A2C1004098.
The work of C.-R.J. was supported in part by the U.S. Department of Energy 
under Grant No. DE-FG02-03ER41260 and in part within the framework of 
the Quark-Gluon Tomography (QGT) Topical Collaboration, under Contract No. DE-SC0023646.
The National Energy Research Scientific Computing Center (NERSC)
 supported by the Office of Science of the U.S. Department of Energy 
under Contract No. DE-AC02-05CH11231 is also acknowledged. }
\end{acknowledgements}

\bibliographystyle{spphys}
\bibliography{ref.bib}
\end{document}